\documentclass[12pt,preprint]{aastex}

\slugcomment{Revised version, submitted to AJ \today}

\shorttitle{First Stellar Abundances in Sextans\,A}
\shortauthors{Kaufer et al.}

\begin{document}
%
%%%%%%%%%%%%%%%%%%%%%%%%%%%%%%%%%%%%%%%%%%%%%%%%%%%%%%%%%%%%%%%%%%%%%%%%%%%
%
\title{First Stellar Abundances in the Dwarf Irregular Galaxy
  Sextans\,A\footnote{Based on UVES observations collected at the
    European Southern Observatory at Paranal, Chile (Proposal IDs
    68.D-0136, 69.D-0383, 70.D-0473).  Additional spectra were
    gathered with the ESI spectrograph at the W.M. Keck Observatory.} }

\author{Andreas Kaufer}
\affil{European Southern Observatory, 
         Alonso de Cordova 3107, 
         Santiago 19, 
         Chile}
\email{akaufer@eso.org}

\author{Kim A. Venn}
\affil{Institute of Astronomy, University of Cambridge,
       Madingley Road, Cambridge, CB3 0HA, UK, {\it and}
       Macalester College, 1600 Grand Avenue, Saint Paul, MN, 55105,
       USA}
\email{venn@macalester.edu}

\author{Eline Tolstoy}
\affil{Kapteyn Institute, 
       University of Groningen, 
       PO Box 800, 
       9700AV Groningen, 
       The Netherlands}
\email{etolstoy@astro.rug.nl}

\author{Christophe Pinte}
\affil{\'Ecole Normale Sup\'erieure, 
       45 rue d'Ulm, 
       F-75005, Paris, 
       France}
\email{christophe.pinte@ens.fr}

\and 

\author{Rolf Peter Kudritzki}
\affil{Institute for Astronomy,
 University of Hawaii at Manoa,
 2680 Woodlawn Drive,
 Honolulu, Hawaii 96822,
 USA}
\email{kud@ifa.hawaii.edu}

%
%%%%%%%%%%%%%%%%%%%%%%%%%%%%%%%%%%%%%%%%%%%%%%%%%%%%%%%%%%%%%%%%%%%%%%%%%%%
%
% TO DO:
%
%%%%%%%%%%%%%%%%%%%%%%%%%%%%%%%%%%%%%%%%%%%%%%%%%%%%%%%%%%%%%%%%%%%%%%%%%%%
%
\begin{abstract}
  
  We present the abundance analyses of three isolated A-type
  supergiant stars in the dwarf irregular galaxy
  \objectname{Sextans\,A} (= DDO\,75) from high-resolution spectra
  obtained with Ultraviolet-Visual Echelle Spectrograph (UVES) on the
  Kueyen telescope (UT2) of the ESO Very Large Telescope (VLT).
  Detailed model atmosphere analyses have been used to determine the
  stellar atmospheric parameters and the elemental abundances of the
  stars.
  The mean iron group abundance was determined from these
  three stars to be
  $\left<[({\rm Fe\,II,Cr\,II})/{\rm H}]\right>=-0.99\pm0.04\pm{\it
    0.06}$\footnote{ In this paper, we use the standard notation
    $[X/{\rm H}]=\log(X/{\rm H})-\log(X/{\rm H})_\odot$. Averages of
    the abundances of the three stars are marked with $\left<\right>$.
    All abundances are given with two uncertainties: first the
    line-to-line scatter, second in italics the estimate of the
    systematic error due to uncertainties in the stellar atmospheric
    parameters.  }.
  This is the first determination of the present-day iron group
  abundances in Sextans\,A.   These three stars now represent the 
  most metal-poor massive stars for which detailed abundance 
  analyses have been carried out.
 
  The mean stellar $\alpha$ element abundance was determined from the
  $\alpha$ element magnesium as 
  $\left<[\alpha({\rm Mg\,I})/{\rm H}]\right>=-1.09\pm0.02\pm{\it
    0.19}$.
  This is in excellent agreement with the nebular $\alpha$ element
  abundances as determined from oxygen in the \ion{H}{2} regions.
  These results are consistent from star-to-star with 
  no significant spatial variations over a length of 
  ~$0.8$\,kpc in Sextans\,A. This supports the nebular abundance
  studies of dwarf irregular galaxies, where homogeneous oxygen 
  abundances are found throughout, and argues against in situ
  (``on the spot'') enrichment.
 
  The $\alpha/{\rm Fe}$ abundance ratio is 
  $\left<[\alpha({\rm Mg\,I})/{\rm
      Fe\,II,Cr\,II}]\right>=-0.11\pm0.02\pm{\it 0.10}$,
  which is slightly lower but consistent with the solar ratio.
  This is consistent with the results from A-supergiant analyses
  in other Local Group dwarf irregular galaxies, NGC\,6822 and WLM.
  The results of near solar [$\alpha$/Fe] ratios in dwarf galaxies is in
  stark contrast with the high [$\alpha$/Fe] results from metal-poor
  stars in the Galaxy (which plateau at values near +0.4~dex), and
  is most clearly seen from these three stars in Sextans\,A because
  of their lower metallicities.
  The low [$\alpha$/Fe] ratios are consistent with the slow chemical 
  evolution expected for dwarf galaxies from analyses of their
  stellar populations.
%, and suggests that the recent star formation
%  epochs in these galaxies have not significantly altered their
%  chemistry, even in situ, from the integrated effects over the 
%  past 15 Gyr.   

\end{abstract}

\keywords{galaxies: abundances, irregular, individual (Sextans\,A) 
          --- stars: abundances}

%
%%%%%%%%%%%%%%%%%%%%%%%%%%%%%%%%%%%%%%%%%%%%%%%%%%%%%%%%%%%%%%%%%%%%%%%%%%%
%
\section{Introduction}

Dwarf irregular galaxies (dIrrs) are low mass, but gas rich galaxies
and are found rather isolated and spread throughout the Local Group.
All dIrrs display low elemental abundances indicating that only little
chemical evolution has taken place over the past 15\,Gyr despite
ongoing star formation. The lack of strong starburst cycles in these
isolated objects may be because of little or no merger interaction.
Therefore, in the context the cold dark matter scenarios of
hierarchical galaxy formation by merger of smaller structures
\citep{Ste99}, dwarf galaxies and in particular the isolated dwarf
irregular galaxies could be the purest remnants of the proto-galactic
fragments from the early Universe.
Hence, the dIrrs are also one of the possible sources for the damped
Ly$\alpha$ absorption (DLA) systems as observed in quasar spectra
over a large range of redshifts; see~e.g.~\citet{Pro03} for a recent
compilation of DLA metallicities over $0.5<z<5$.
%
%A thorough examination of 
The dIrrs being possible remnants of the early Universe might
allow us to study early galaxy evolution in great detail in the nearby
universe, even in the Local Group.  In particular the early chemical
evolution of galaxies is of interest here to shed light on the
formation of the first generations of stars.
The most powerful way to study the chemical evolution of a galaxy is
via the elemental abundances and abundance ratios of their stellar
and gas content which contain a record of the star formation histories
(SFH) of the galaxy over the last 15\,Gyr.

The analysis of bright nebular emission lines of \ion{H}{2} regions
has been the most frequent approach to modeling chemical evolution of
more distant galaxies to date \citep{Mat85}. So far, only a very
limited number of elements can be examined and quantified when using
this approach. The chemical evolution of a galaxy depends on the
contributions of all its constituents, e.g., SNe type Ia and II, high
mass stars, thermal pulsing in low and intermediate mass AGB stars.
Thus, more elements than just those observed in nebulae need
to be measured, since each have different formation sites which sample
different constituents. The $\alpha$ elements like oxygen are created
primarily in short-lived massive stars, while the iron-group elements
are produced in SNe of both high and low mass stars. Therefore, it is
expected that substantially different SFHs of individual galaxies
shall be represented in different present-day $\alpha/{\rm Fe}$ ratios
as discussed e.g. in \citet{Gil91}. In reverse, the accurate
measurement of the absolute stellar abundances and abundance ratios
allows to put constraints on the SFH of the respective galaxy.
In addition to the study of the average metallicities of a galaxy as
discussed above, also the spatial distribution of abundances and
abundance patterns within a galaxy contains important information
about the redistribution processes of the newly produced elements and
their subsequent mixing with the ISM. If the abundances of a number of
stars distributed over the galaxy can be analyzed, possible chemical
inhomogeneities e.g.~in form of large spreads in the abundances or
abundance gradients can be studied and provide additional constraints
of the chemo-dynamical evolution of the galaxy.

Unfortunately, most dIrr systems are too distant for the detailed
study and abundance analyses of e.g.~their red giant branch (RGB)
stars, which are of great importance due to their large spread in age.
Hence, to date few details on the chemical evolution of the dIrr
galaxies have been available. In the case of the dIrr Sextans\,A
studied in this work the tip of the RGB is found at visual magnitudes
of $V\approx23$ \citep{DP02} --- far beyond the capabilities of even
the most efficient high-resolution spectrographs on today's 8 to
10-meter class telescopes.
However, with the same class of latest telescopes and instrumentation,
the visually brightest stars of the dIrr galaxies, i.e., the blue
supergiants ($17.5\lesssim V\lesssim 20$ in Sextans\,A) become now
accessible to detailed spectroscopy, which allow us to
determine the present-day abundances of $\alpha$ {\it and} iron group
elements. The abundances of the $\alpha$ elements derived from the
rich spectra of hot massive stars can be directly compared and tied to
the nebular $\alpha$ element abundances because both nebulae and
massive stars have comparable ages and the same formation sites.
The first abundance studies based on high-resolution spectra of 
A-type supergiant stars were carried out for the nearby Local Group
dIrrs NGC\,6822 and WLM \citep{Ven01,Ven03a}. 
$\alpha$ element and iron group abundances
could be derived for two stars in both dIrr galaxies.
For NGC\,6822 and WLM the [$\alpha/{\rm Fe}$] ratios are in good 
agreement with the {\it solar} ratio.    
This result is somewhat surprising since both galaxies have ongoing
star formation and different star formation histories as determined
from their stellar populations (e.g., \citet{Gal96}, \citet{Mat98}, 
\citet{Dol03a}).

The isolated dwarf irregular galaxy Sextans\,A (= DDO\,75) with its
distance of about $1.3$\,Mpc \citep{Dol03a} is located at a similar
distance as Sextans\,B, NGC\,3109, and the Antlia dwarf galaxies which
form a small group of galaxies at the edge of the Local Group ---
possibly not even bound to the Local Group which would make Sextans\,A
a member of the nearest (sub)cluster of galaxies \citep{vdB99}.
The present-day chemical composition was studied from the
emission-line spectroscopy of bright compact \ion{H}{2} regions
\citep{Hod94}. The analysis by \citet{Ski89} determines an oxygen
abundance of $12 + \log(\rm{O/H}) = 7.49$ which corresponds to
$[\rm{O/H}] =-1.17$ if a solar oxygen abundance of $12 +
\log(\rm{O/H})_\odot = 8.66$ is adopted \citep{Asp03}. In a revised
analyses of the same data by \citet{Pil01}, a higher value of
$12 + \log(\rm{O/H}) = 7.71$ is found, 
corresponding to $[\rm{O/H}] =-0.95$.
The star formation history of Sextans\,A was first studied in detail
by \citet{DP97}, and more recently by \citet{DP02} and \citet{Dol03b}
using HST VI color -- magnitude diagrams (CMD). 
A high rate of star formation has occured in the past $~2.5$\,Gyr,
with very little star formation between $2.5$ and $10$\,Gyrs. 
A mean metallicity of $[\rm{M/H}]\approx-1.45\pm0.2$ is derived over 
the measured history of the galaxy, with most of the enrichment 
having occurred more than 10\,Gyr ago.   Only slight enrichments
(+0.4\,dex) occured between 2 and 10 Gyr from fitting the red giant
branch to a metallicity of $[\rm{M/H}]\approx-1.1$.   

In this paper, we present the results from a detailed abundance
analysis of three isolated and normal A supergiant stars in Sextans\,A
based on high resolution spectra obtained at the ESO-VLT using UVES.
We compare the stellar results to those from nebular analyses 
to examine the chemical homogeneity of this young galaxy.  We also
use the rich stellar spectra to compare the abundances of several
other elements not available in the nebular analyses to those
predicted from its chemical evolution history, as well as to results
from other dwarf galaxies.

%%%%%%%%%%%%%%%%%%%%%%%%%%%%%%%%%%%%%%%%%%%%%%%%%%%%%%%%%%%%%%%%%%%%%%%%%%%
%
\section{Observations}

%
% target selection
% 
A list of potential targets for high-resolution UVES \citep{Dek00}
spectroscopy at the ESO VLT was selected from the photometric
catalogue by \citet{vDyk98}. The visually brightest ($V<19.8$) stars
with colors ranging from $-0.10<(B-V)<0.30$ were chosen from the
corresponding CMD.
The selected targets were checked for possible crowding on HST/WFPC2 
images from \citet{DP97}.
For one of the targets, the star number 754 in the catalogue of
\citet{vDyk98} (in the following identified as ``SexA-754''), a low
resolution Keck/LRIS \citep{Oke95} spectrum was obtained by
\citet{Kob98}, which confirmed that this star is a member of
Sextans\,A (cf. below) and a normal A supergiant suited for abundance
analyses.
Almost 14 hours of VLT/UVES spectroscopy of SexA-754 were carried out
in service mode in between January and February 2002 within the
specified constraints of a seeing better than $1.2$\arcsec, a dark and
clear sky, and an airmass $<1.5$. Additional spectra in complementary
wavelength settings in the red were obtained in a visitor mode run in
April 2002 by AK,KAV,CP.
In parallel Keck/ESI \citep{She02} low resolution spectroscopy of
other potential targets was obtained by KAV in February 2002. 
For the best candidates, the ESI low-resolution spectra were
complemented by high-resolution spectra taken with UVES in the April
2002 visitor run.
%
% selection process
%
All candidates stars were first checked for their membership to the
Sextans\,A galaxy by comparing the measured radial velocities of the
stars to the heliocentric radial velocity of Sextans\,A as determined
from optical and radio spectroscopy: \citet{Tom93} determined values of
$325-335$\,km/s from the H$\alpha$ emission lines of different
\ion{H}{2} regions, which are consistent with earlier measurements of
the radial velocities of the \ion{H}{1} gas, i.e., $324\pm3$
\citep{Huc86} and $325\pm5$\,km/s \citep{Ski88}. Due to the large
heliocentric radial velocity of Sextans\,A, any confusion of our
stellar targets with foreground stars is very unlikely. 

Based on the collected ESI and UVES spectra two additional stars
(SexA-1344 and SexA-1911) were identified for 2 times 12.5 hours of
VLT/UVES service mode observations, which were executed between
December 2002 and February 2003 within the same constraints as given
for SexA-754 before. The candidate stars SexA-513, SexA-983, and
SexA-1456 on the other hand were found to be not suited for our
abundances analyses which are tailored for A supergiants of a narrow
temperature and luminosity range and the three stars had to be
discarded.

%
% Obs log and sample
%
A complete log of the ESI and UVES observations is found in
Tab.~\ref{SexA-observations} together with some additional information
on the instrumental configurations used for the individual observations
and the environmental conditions during the actual observations.
The complete sample of examined Sextans\,A targets is presented in
Tab.~\ref{SexA-sample} together with the radial velocity and spectral
type information as obtained either during the detailed analyses of
the stars or for the discarded stars during the pre-selection process;
the location of the individual targets within the galaxy field is
shown in Fig.~\ref{SexA-field} together with the positions of the four
bright compact \ion{H}{2} regions analyzed by \citet{Ski89}.

%
% Data reduction
%
All UVES data were reduced using the dedicated UVES pipeline which is
described by \citet{Bal00} and is available as a special package
within the ESO-MIDAS\footnote{Munich Image Data Analysis System} data
reduction system. Optimum extraction with cosmic rejection and sky
subtraction was used to extract the individual stellar spectra.  A
typical signal-to-noise ratio (SNR) of $\approx9$ in the wavelength
regime corresponding to the $B$ band was obtained for a single
spectrum of 4500\,sec exposure time.
The typically 10 individual spectra per star were then corrected for
the spectrograph's response function to obtain a smooth continuum, 
rebinned to heliocentric velocities, and eventually combined with 
additional cosmic clipping.
The combined spectra were resampled and rebinned to a 2-pix resolving
power of $R=20\,000$ which was used throughout the analyses.  The
reduced resolving power allows to increase the SNR per resolution
element without having to sacrifice line profile information. The
projected rotational velocities of the three stars were later
determined to be $>30$\,km/s, i.e., all line profiles remain fully
resolved at the reduced resolution of $15$\,km/s. The SNRs obtained for
the combined UVES spectra are listed in Tab.~\ref{SexA-snr} for
different wavelength regions of interest.
The observations of individual stars in a distant Local Group galaxy 
as presented here are at the limit of the capabilities of today's largest
telescopes and most sensitive spectrographs: despite total integration 
times of typically 12.5\,hrs per star, SNRs of 30 are achieved at best
for a resolving power of $R=20\,000$ for a $V=19.5$\,mag star which is
deemed just sufficient for detailed abundances analyses as will be 
shown in this work. 

In Fig.~\ref{SexA-Halpha} the H$\alpha$ profiles of the three stars
are shown. To obtain the stellar line profiles, the contribution of
the diffuse H$\alpha$ emission of Sextans\,A which was found for all
stars was subtracted as part of the 'sky' subtraction in the
extraction process of the spectra. Inspecting the raw 2D spectra, the
diffuse H$\alpha$ emission does show spatial structure but averaged
along the slit its distribution is mostly Gaussian. The diffuse
emission is blueshifted by an average of $-10$\,km/s with respect to
the stellar spectra and consistent with the \ion{H}{1} radial velocity
of Sextans\,A \citep{Huc86}. The average FWHM of the emission is
$47$\,km/s corresponding to a ($1\sigma$) velocity dispersion of
$20$\,km/s. This is in excellent agreement with a thermal velocity
($22$\,km/s) of a warm interstellar medium if a temperature of
$10\,000$\,K is assumed.

In the sky-subtracted stellar H$\alpha$ profiles in
Fig.~\ref{SexA-Halpha} no stellar line-emission is visible in
H$\alpha$ which indicates that the selected stars are supergiants of
lower luminosity.  This was already expected from the fact that during
the selection process only stars about 2 magnitudes fainter than the
brightest blue stars in Sextans\,A (SexA-513) turned out to be normal
A supergiants.  Interestingly, no brighter, more extreme A
supergiants, which are supposed to be the visually brightest stars,
were found in the whole galaxy. The same observation was made during
the study of the dIrr WLM \citep{Ven03a}. The selection process
applied to find candidate stars for abundance analyses in Sextans\,A
and WLM was not rigorous enough to exclude that this finding is a
selection effect intrinsic to our procedure.  However, the lack of
luminous A supergiants might as well be an evolution effect due to the
low metallicity of the galaxy.
Similarly, the lack of wind emission features in H$\alpha$ as seen
in Fig.~\ref{SexA-Halpha} might be due to the low metallicity 
suppressing the formation of stellar winds due to the reduced 
radiation line-driving on the metal-poor gas.

Lower-luminosity supergiants as were found in Sextans\,A are best
suited for abundance analyes.  Higher-luminosity stars are difficult
to analyze because the line-forming regions of their atmospheres are
no longer in hydrostatic equilibrium but are found in the acceleration
zone of the stellar wind.
Further, for lower-luminosity A supergiants, the effects due to
intrinsic photospheric and wind variability on the line profiles is
negligible \citep{Kau96}.  Equivalent width variations of weak
($W_\lambda<200$\,m\AA) photospheric lines as used for the abundance
analyses in this work are found to be $\le10$\,\% \citep{Kau97}.

%

%
%%%%%%%%%%%%%%%%%%%%%%%%%%%%%%%%%%%%%%%%%%%%%%%%%%%%%%%%%%%%%%%%%%%%%%%%%%%
%
\section{Atmosphere Analyses}

The atmosphere analyses of the three stars in Sextans\,A have been
carried out in line with previous analyses of A supergiant stars in
the Galaxy \citep{Ven95a,Ven95b}, the Magellanic Clouds \citep{Ven99},
M31 \citep{Ven01} and the dwarf irregular galaxies NGC\,6822
\citep{Ven01} and WLM \citep{Ven03a}. Maintaining a consistent
analysis process across the different studies of A supergiant stars is
considered crucial to obtain consistent and comparable abundance
results. In particular results from differential analyses can then be
considered the most reliable because the uncertainties in the
determined absolute elemental abundances shall cancel out at least to
some extend.

%
% Atlas atmospheres
%
The atmosphere computations are based on plane parallel, hydrostatic,
and line-blanketed ATLAS\,9 LTE model atmospheres
\citep{Kur79,Kur88,Kur93} which are considered to be appropriate for
the photosphere analysis of lower luminosity A supergiants
\citep{Prz02}.
To account for the possible effects of the low metallicity of
Sextans\,A on the model atmosphere structure, all atmosphere models
have been computed with a metallicity scaled to $1/10$ of solar and a
corresponding opacity distribution function. The value of $[{\rm
  M}/{\rm H}]=-1$ was chosen in agreement with the present day oxygen
abundances as derived from the analysis of \ion{H}{2} regions
\citep{Ski89}.  However, as has been shown later, the effect of
atmosphere models with different metallicities on the derived
elemental abundances is typically less than $0.05$\,dex, i.e.,
negligible within the estimated systematic errors.

%
% weak lines
%
The analyses has been restricted to the use of weak absorption lines
only to minimize the effects of the neglected spherical extension and
of the neglected NLTE conditions in these extended, low gravity 
supergiant stars on the model atmospheres. Further, weak lines are
preferred because they suffer less from NLTE and microturbulence
effects in the line formation process.
Weak lines in this context are defined as lines where a change in 
the microturbulence $\xi$ of $\Delta\xi=1$\,km/s results in a change
in the elemental abundance of $\log(X/\rm{H})\le0.1$\,dex. To fulfill
this weak-line condition, the equivalent width of the corresponding
line has typically to be $W_\lambda<160$\,m\AA. 
Weak-line analyses are a challenge for faint and metal-poor targets 
as analyzed in this work, in particular if the galaxies are far
away, the targets are faint and the SNR of the high-resolution
spectra is comparatively low as it is the case here.
The absorption lines have been measured by fitting Gaussian profiles
to the weak line profiles and to the stellar continuum, the latter
being the main source of uncertainty in the derived equivalent widths
$W_\lambda$. The comparatively low SNR leaves a non-negligible
uncertainty of about $10$\% in the definition of the continuum level
and therefore in the measured $W_\lambda$ in the blue part of the
combined spectra.

The stellar model atmosphere parameters effective temperature $T_{\rm
  eff}$, gravity $\log g$, and microturbulence $\xi$ are in the
following determined solely from spectral features and therefore are
not affected by the (badly-defined) extinction.

%
% teff - logg determination
%
The temperature and gravity parameters of the best fits of synthetic 
Balmer-line profiles to the gravity-sensitive wings of the H$\gamma$ 
profiles (see Fig.~\ref{SexA-Hgamma}) define a relation in the 
$T_{\rm eff}$ -- $\log g$ plane as shown for each star in 
Fig.~\ref{SexA-tefflogg}. 
The temperature and gravity parameters for which an ionization 
equilibrium of \ion{Fe}{1} and \ion{Fe}{2} is found defines a 
temperature-sensitive relation in the $T_{\rm eff}$ -- $\log g$ plane
(for a detailed discussion of the \ion{Fe}{1} NLTE corrections see 
below).
The intersection of both relations defines the position in the 
$T_{\rm eff}$ -- $\log g$ plane for which both, the best fit to
the H$\gamma$ profile and the ionization equilibrium of \ion{Fe}{1} 
and \ion{Fe}{2} are achieved. This parameter pair is adopted as
the atmosphere parameters of the star.

%
% FeI/II equilibrium
%
Due to the low temperature of the analyzed stars and the comparatively
low SNR of the combined spectra (cf.~Tab.~\ref{SexA-snr}) none of the
weak spectral lines of \ion{Mg}{2} present in the observed spectral
ranges could be used to determine the stellar parameters from the most
reliable ionization equilibrium of \ion{Mg}{1} and \ion{Mg}{2}.  Also
the strong, easy to measure \ion{Mg}{2}$\lambda4481$ line had to be
discarded because of its known high sensitivity to NLTE and
microturbulence effects: NLTE corrections are of the order of
$-0.3$\,dex for the temperature and gravities of the stars analyzed
here; an uncertainty of the microturbulence of $1$\,km/s results in a
change of the computed abundances of the same order. Instead, as
already described above, the ionization equilibrium of \ion{Fe}{1} and
\ion{Fe}{2} was used in this work.  This introduces an additional
uncertainty to the determination of the stellar parameters because of the
sensitivity of \ion{Fe}{1} lines to NLTE effects \citep{Boy85, Gig86,
  RH96}. The NLTE effects on \ion{Fe}{2} lines are considered
negligible \citep{RH96,Bec98}.
%
%
% FeI NLTE corrections
%
To quantitatively incorporate the dependencies of the \ion{Fe}{1} NLTE
corrections on temperature, gravity, and metallicity on the
\ion{Fe}{1}/\ion{Fe}{2} ionization equilibrium, the results from
\citet{RH96} presented in her Figs.~4 and 5 were used to derive
a relation for the NLTE correction to be applied to the \ion{Fe}{1}
abundances for the limited temperature range of interest in this work,
i.e., $7000<T_{\rm eff}<9000$\,K:
\begin{eqnarray}
\label{eqnFeINLTE}
\Delta\log\epsilon_{\rm Fe I}
      &=& 0.09 \frac{T_{\rm eff}[K]}{1000}-0.58              \nonumber\\ 
      & & -2(\log g - 4) 
          \left( 0.04\frac{T_{\rm eff}[K]}{1000}-0.27\right) \nonumber\\ 
      & & -0.1 [{\rm M}/{\rm H}]                             \nonumber\\ 
      & & -{\it 0.28}                                        \nonumber\\ 
\end{eqnarray}
The first and second terms describe the temperature and gravity
dependent NLTE correction as derived for solar metallicity main
sequence ($\log g \approx 4.0$) stars. The third term describes the
metallicity dependence of the NLTE correction as derived for
metallicities in the range $[{\rm M}/{\rm H}]=\pm0.5$. Due to the
considerably lower gravities and metallicities of the stars studied
here, this relation has to be used with caution.

To test the relation for stars of lower gravity and metallicity as
studied in this work the relation was applied to the data set of A
supergiants in the SMC as analyzed by \citet{Ven99}. The stellar
parameters of the SMC stars are all based on the most reliable 
\ion{Mg}{1} to \ion{Mg}{2} ionization equilibrium. The \ion{Fe}{1}
NLTE corrections were determined as $\Delta\log\epsilon_{\rm
  Fe\,I}=\log({\rm Fe\,II/H})-\log({\rm Fe\,I/H})$ assuming that the NLTE
corrections for \ion{Fe}{2} are negligible \citep{Bec98,RH96}.  It is
found that the above relation fits the SMC data well if an additional
correction of $-0.28$ is added, cf.~Fig.~\ref{SexA-FeINLTE}. It is
important to note that the relation describes rather well all SMC stars
with \ion{Fe}{1} to \ion{Fe}{2} abundance differences over the whole
range from $0.0$ to $0.3$\,dex.
This constant additional (fourth) term in the relation seems
sufficient to absorb the additional NLTE effects due to low gravity
and low metallicity, at least for the narrow temperature range of
$7000-9000$\,K, gravities of $\log g\approx1.5$, and metallicities of
$[{\rm M}/{\rm H}]\approx-0.7$, i.e., close to the stellar parameters
of the stars studied in Sextans\,A. In the subsequent atmosphere
analysis, the final stellar parameters for the Sextans\,A stars were
chosen to be consistent with the \ion{Fe}{1} NLTE corrections as given
by the above relation.  The adopted final corrections are $+0.23$\,dex
for SexA-754, $+0.30$\,dex for SexA-1344, and $+0.20$\,dex for
SexA-1911 and are shown in Fig.~\ref{SexA-FeINLTE}, too.
It should be further noted that the remaining uncertainty of
$\pm0.1$\,dex in the \ion{Fe}{1} NLTE corrections has been taken into
account in the overall error budget as estimated from the
uncertainties of the determination of the stellar parameters
(cf.~below).

%
% Microturbulence
%
Microturbulence $\xi$ has been determined from the line abundances of
the numerous lines of different strength from the ions \ion{Fe}{2},
\ion{Ti}{2}, and \ion{Cr}{2}. A single value of $\xi$ for each star was
adopted so that no trend of the computed elemental abundance is found
with the equivalent width of the lines within the uncertainty of 
$\Delta\xi=1$\,km/s. 
%
% Error estimates
%
%Uncertainties in $T_{\rm eff}$ are estimated from the uncertainties of
%the \ion{Fe}{1} NLTE correction of $\pm0.1$\,dex to be $\Delta T_{\rm
%  eff}=\pm200$\,K. The effect of a change of the \ion{Fe}{1} NLTE
%correction of $\pm0.08$\,dex is shown in Fig.~\ref{SexA-tefflogg} for
%star SexA-754. 
%
The estimated uncertainties in $T_{\rm eff}$ are dominated by the
uncertainties of the derived \ion{Fe}{1} NLTE corrections but also
have to account for the uncertainties of the assumption of negligible
NLTE effects on \ion{Fe}{2} in the ionization equilibrium used for the
temperature determination. The effect of a change of the \ion{Fe}{1}
NLTE correction of $\pm0.08$\,dex (as derived from the residuals
in Fig.~\ref{SexA-FeINLTE}) is shown in Fig.~\ref{SexA-tefflogg}
for star SexA-754. If we assume an uncertainty of $\pm0.05$\,dex for
the LTE assumption for \ion{Fe}{2}, we obtain an estimate for the
total uncertainty of $\pm0.1$\,dex which translates into $\Delta
T_{\rm eff}=\pm200$\,K as indicated by the horizontal errorbars
in Fig.~\ref{SexA-tefflogg}.
Uncertainties in gravity are estimated to $\Delta\log
g=\pm0.2$\,dex from varying the fits to H$\gamma$ for different $\log
g$ at constant $T_{\rm eff}$. The resulting change of the
synthetic profile for a variation of $\Delta\log g = +0.2$\,dex is
shown in Fig.~\ref{SexA-Hgamma}.
%
% vsini
%
With the determined stellar parameters and elemental abundances (cf.
below), a small section of the spectrum between $4285-4305$\,\AA\ was
synthesized to determine the projected rotational velocities $v\sin i$
of the three stars. This spectral region contain several stronger and
isolated \ion{Ti}{2} and \ion{Fe}{2} lines.  For a given resolving
power of $R=20\,000$ or $15$\,km/s and the given heliocentric radial
velocities of the stars the respective values of $v\sin i$ were
obtained by least-square fitting of the synthetic line profiles to the
observed spectra while keeping the elemental abundances for the metal
lines fixed to the value determined before. The resulting fits are
shown in Fig.~\ref{SexA-vsini} and the corresponding values for $v\sin
i$ are reported in Tab.~\ref{SexA-analysis}.

%
% Spectral type, Luminosity, Radius
%
The spectral types of the three stars were defined according to
their effective temperatures according to \citet{SK82}. It should
be noted that this spectral classification does not take into account
any metallicity dependence of the classification criteria but is
based on stars of solar metallicities.
The luminosities of the three stars were then derived from the visual
magnitudes as given in Tab.~\ref{SexA-sample} using the distance
modulus of $\mu_0=25.6$ from \citet{Dol03a} and the (close to zero)
bolometric corrections from \citet{SK82} corresponding to the spectral
types.
The resulting spectral types and luminosities are given in
Tab.~\ref{SexA-analysis} together with the stellar radii as computed
from the luminosity and the effective temperature. The luminosities of
all three stars are consistent with the evolutionary track of a
$12\,{\rm M}_\odot$ star \citep{Lej01}, i.e., a ZAMS mass compatible
with evolved low-luminosity A supergiants. This finding further
confirms the Sextans\,A membership of the three analyzed supergiant
stars.

%
%%%%%%%%%%%%%%%%%%%%%%%%%%%%%%%%%%%%%%%%%%%%%%%%%%%%%%%%%%%%%%%%%%%%%%%%%%%
%
\section{Abundances}
\label{abunds}

The list of appropriate spectral lines for the abundance
determinations and the corresponding atomic data were adopted from the
previous analyses of \citet{Ven95a,Ven95b,Ven99,Ven01}. Atomic data
and the abundances derived from the individual lines are listed in
Tab.~\ref{lines} together with the measured equivalent widths 
$W_\lambda$.
All abundance calculations in LTE/NLTE and LTE spectrum synthesis 
have been carried out using a modified version of the LINFOR code\footnote{
  LINFOR was developed by H.~Holweger, W.~Steffen, and W.~Steenbock
  at Kiel University. Later modifications were added by M.~Lembke and
  N.~Przybilla. The code is available at \anchor{http://www.sternwarte.uni-erlangen.de/pub/MICHAEL/ATMOS-LINFOR-NLTE.TGZ}{\url{http://www.sternwarte.uni-erlangen.de/pub/MICHAEL/ATMOS-LINFOR-NLTE.TGZ}}.
}.
%
% Abundance results
%
The resulting average elemental abundances are given for each
ionization stage in Tab.~\ref{SexA-analysis}. For each abundance, two
error estimates are given, first the r.m.s. line-to-line scatter,
second (in italic) the estimated systematic error in the elemental
abundance from the uncertainties in the atmospheric parameters
$T_{\rm eff}$, $\log g$, and $\xi$.
The systematic errors of the average abundances have been estimated
by the variation of the individual stellar atmosphere parameters
within their estimated uncertainties while keeping the other
parameters fixed. The individual error estimates are reported in
Tab.~\ref{SexA-abuerror}. From the error estimates it can be seen that
\ion{Si}{2}, \ion{Ti}{2}, \ion{Cr}{2}, and \ion{Fe}{2} provide the
most reliable elemental abundances. 
However, abundances derived from the \ion{Si}{2} ion have been found
to suffer from a large spread if derived from stars over a wider range
in stellar parameters. The sample of 10 A supergiants in the SMC as
analyzed by \cite{Ven99} shows a spread in the silicon abundances of
$\approx0.5$\,dex probably due to neglected NLTE effects in the line
formation of the \ion{Si}{2} ion.  The three stars analyzed here show
very consistent silicon abundances because the stellar parameters of
the stars are very close to each other. However, since the absolute
abundance probably is affected from the uncertainties that were found
in other analyses, the abundances from \ion{Si}{2} will not be
considered in the further discussion of $\alpha$ element abundances.

The average elemental abundances relative to the solar abundances are
shown for the three stars in Fig.~\ref{SexA-eps2solar} with the error
bars representing the estimated combined systematic errors in the
abundances. The combined errors were computed as the quadratic sum of
the individual errors from Tab.~\ref{SexA-abuerror}.

%
% alpha elements
%
%
% Mg NLTE
%
The abundances reported for \ion{Mg}{1} and \ion{Mg}{2} have been
computed using NLTE lineformation based on the model atom by
\citet{Gig88} which delivers results in good agreement with the model
atom developed by \citet{Prz01}. The NLTE corrections for the
individual Mg lines are listed in Tab.~\ref{SexA-Mg_corr} and are
found to be small and of the order of $\pm0.1$\,dex.
Abundances from the strong \ion{Mg}{2}$\lambda4481$ line blend were
derived from fits to the observed line profile using spectrum
synthesis in LTE; the results are reported in Tab.~\ref{lines}. The
NLTE corrections for this line are expected to be large \citep{Prz01}
and very sensitive to the stellar parameters and in particular to the
microturbulence \citep{Prz03}. 
Therefore, no attempt was made to compute the NLTE
corrections and the line was not used for this analysis but is given
for reference only.

For star SexA-754, useful upper limits for the equivalent widths
of the weak \ion{Mg}{2} lines could be given because of the higher SNR
of the combined spectrum of this star. The resulting upper limits for
the \ion{Mg}{2} abundances are consistent with the abundances derived
from the \ion{Mg}{1} lines and therefore further support the stellar
parameters as derived from the \ion{Fe}{1} to \ion{Fe}{2} ionization
equilibrium.

%
% Oxygen I, SexA-754
%
The $\alpha$ element of highest interest is oxygen. Unfortunately,
none of the \ion{O}{1} lines in the 615\,nm and 645\,nm spectral
regions could be measured in the combined UVES spectra of moderate SNR
because of the weakness of the lines at the metallicity of Sextans\,A.
For $[{\rm O/H}]\approx-1$, all \ion{O}{1} lines in this spectral
region are expected to have equivalent widths of $<20$\,m\AA, i.e.,
clearly below the detection limit of the spectra.
However, the low metallicity of Sextans\,A brings the near-infrared
\ion{O}{1}$\lambda\lambda7771$-$7775,8446$ lines of multiplet 1 and 4
into the weak-line regime possibly suited for abundances analyses
despite their known high sensitivity to NLTE effects \citep{Prz00}. 

Spectra of the respective wavelength regions are only available for
one of the stars, SexA-754.  Since the near-infrared lines are
non-resolved multiplets, fitting of synthetic LTE line profiles was
used to determine LTE oxygen abundances (see Fig.~\ref{SexA-754_OI}).
The oxygen abundances measured in LTE are $12+\log{\rm{O/H}}=9.1\pm0.3$ 
for \ion{O}{1}$\lambda7771$-$7775$ and $12+\log{\rm{O/H}}=8.1\pm0.3$ for
\ion{O}{1}$\lambda8446$. 
Detailed computations of the NLTE corrections for the given stellar
parameters of SexA-754 at the metallicity of Sextans\,A will be
required before accurate oxygen abundances can be derived for this
star. In any case, the expected uncertainties will remain large, first
because of the large required deviations from LTE to bring the oxygen
abundances into the reasonable range of $12+\log{\rm{O/H}}\approx7.8$
(assuming an underabundance of $[\alpha/{\rm H}]=-1.0$ as suggested from
the other $\alpha$ elements silicon and magnesium), second because of
the low quality of the near-infrared spectra, which results in a
non-negligible uncertainty in the continuum definition, which is
crucial for an accurate line profile modeling. The estimated error in
the derived LTE abundances of $\pm0.3$\,dex is due to this uncertainty in
the continuum localization.
Therefore, currently no further attempt has been made to determine the
NLTE corrections for the near-infrared lines of SexA-754. In the
following, elemental abundances derived from \ion{Mg}{1} will be used
instead for the discussion of $\alpha$ element abundances in
Sextans\,A.
Earlier studies in the SMC \citep{Ven99} and NGC\,6822 \citep{Ven01}
have shown that a good agreement between the $\alpha$ elements oxygen
and magnesium is found for different stars and so both elements may be 
used to represent the $\alpha$ elemental abundance of the star. 
The average $\alpha$ element abundance of the three stars in
Sextans\,A are found to be 
$\left<[{\rm Mg\,I/H}]\right>=-1.09\pm0.02\pm{\it 0.19}$ 
which is in excellent agreement with the
$\alpha$ element abundances from the \ion{H}{2} regions of $[\rm{O/H}]
=-1.17$ as derived by \citet{Ski89} and the recently revised value
$[\rm{O/H}] =-0.95$ of \citet{Pil01}.

%
% iron group elements
%
The iron-group elemental abundances of the three stars are in
excellent agreement as derived from lines of \ion{Cr}{1}, \ion{Cr}{2},
\ion{Fe}{1}, and \ion{Fe}{2}. The good agreement of the abundances
from the two different ionization stages of chromium is a further
indication for correct stellar parameter. However, it should be noted
that this equilibrium is based on a single \ion{Cr}{1} line. Possible
NLTE effects on this line which were not taken into account in the
abundance computation appear therefore to be small. The agreement of
the \ion{Fe}{1} and \ion{Fe}{2} abundances and hence the ionization
equilibrium is achieved if the respective \ion{Fe}{1} NLTE corrections
are applied, i.e., $+0.23$\,dex for SexA-754, $+0.30$\,dex for
SexA-1344, and $+0.20$\,dex for SexA-1911. As discussed before, the
determination of the stellar parameter is based specifically on this
NLTE corrected ionization equilibrium.
The abundances of \ion{Fe}{2} and \ion{Cr}{2} have been derived from
numerous lines and are regarded as the most reliable: the dependency
of the abundances on the stellar parameters is small
(cf.~Tab.~\ref{SexA-abuerror}) and possible NLTE effects on these
dominant ionization stages are expected to be small.

The mean metallicity of the three stars are so found to be
$\left<[({\rm Fe\,II,Cr\,II})/{\rm H}]\right>=-0.99\pm0.04\pm{\it 0.06}$
which --- to our knowledge --- makes these three stars the to date most
metal-poor massive stars in any galaxy for which a detailed abundance
analysis has been carried out.

%
% Ti and Sc: alpha or iron group
%
Titanium and scandium can be considered as either $\alpha$ or iron-group
elements.
The Ti abundances were derived from a large number of lines 
of the dominant ionization stage of singly ionized titanium and are 
in good agreement with iron, i.e.,
$\left<[{\rm Ti\,II}/{\rm Fe\,II}]\right>=0.08\pm0.05\pm{\it 0.08}$
while Sc shows a slight underabundance of
$\left<[{\rm Sc\,II}/{\rm Fe\,II}]\right>=-0.20\pm0.03\pm{\it 0.09}$. 
This might be due to the smaller number of available lines and higher
temperature sensitivity of \ion{Sc}{2}. Furthermore, the
hyperfine-structure terms in the lines of odd elements of the
iron group due the presence of a nuclear magnetic moment have been
neglected.  Due to these uncertainties, no further discussion of the
scandium abundances will follow.

%Abundances averaged for the three stars:
%$\alpha$ elements:\\
%$\left<[{\rm Mg\,I/H}]\right> =-1.09\pm0.02\pm{\it 0.18}$ \\
%$\left<[\alpha({\rm Mg\,I})/{\rm H}]\right>=-1.09\pm0.02\pm{\it 0.19}$ 

%$\alpha$ and iron group elements:\\
%$\left<[{\rm Ti\,II/H}]\right>=-0.96\pm0.09\pm{\it 0.12}$ \\
%$\left<[{\rm Sc\,II/H}]\right>=-1.24\pm0.03\pm{\it 0.14}$ \\

%Iron group elements:\\
%$\left<[{\rm Cr\,I/H}]\right>=-0.94\pm0.05\pm{\it 0.19}$  \\
%$\left<[{\rm Cr\,II/H}]\right>=-0.94\pm0.06\pm{\it 0.07}$ \\
%$\left<[{\rm Fe\.I/H}]\right>=-1.03\pm0.05\pm{\it 0.19}$  \\
%$\left<[{\rm Fe\,II/H}]\right>=-1.03\pm0.05\pm{\it 0.11}$ 

%$\left<[({\rm Fe\,II,Cr\,II})/{\rm H}]\right>=-0.99\pm0.04\pm{\it 0.06}$

%$\alpha/{\rm Fe}$:\\
%$\left<[\alpha({\rm Mg\,I})/{\rm Fe\,II}]\right>=-0.06\pm0.03\pm{\it 0.11}$ 
%$\left<[\alpha({\rm Mg\,I})/{\rm Fe\,II,Cr\,II}]\right>=-0.11\pm0.02\pm{\it 0.10}$ 

%
%%%%%%%%%%%%%%%%%%%%%%%%%%%%%%%%%%%%%%%%%%%%%%%%%%%%%%%%%%%%%%%%%%%%%%%%%%%
%
\section{Discussion}

%
% present day chemistry of SextansA
%

For the first time, the present-day iron group abundances have been
determined from stars in Sextans\,A. The mean underabundance relative
to solar is 
$\left<[{\rm Fe/H}]\right>=-1.03\pm0.05\pm{\it 0.11}$ 
and has been derived from the \ion{Fe}{2} abundances which are
expected to be the most reliable. This result is well supported by
abundances measured for a second iron-group element, namely
\ion{Cr}{2} with a mean underabundance of
$\left<[{\rm Cr\,II/H}]\right>=-0.94\pm0.06\pm{\it 0.07}$.
The individual iron-group abundances of the three stars from which the
mean abundances were computed are in good agreement within the
estimated systematic errors.
%
%However, a small overabundance of
%$\approx0.25$\,dex for Fe and $\approx0.20$\,dex for Cr of star 
%SexA-1344 with respect to SexA-754 and SexA-1911 could be derived. 
%This result is probably not significant --- an adjustment of the stellar 
%parameters by about $400$\,K in effective temperature would bring the 
%iron abundances in full agreement.

The star formation history of Sextans\,A derived by \citet[and
references therein]{Dol03b}, suggests that the metallicity reached
$[{\rm M}/{\rm H}]=-1.1 \pm0.2$ only 2 Gyr ago.  The metallicity
spread of $\pm 0.2$\,dex is to account for the dispersion in the Blue
Helium Burning (BHeB) stars.
The mean stellar metallicity found here from iron and chromium, 
$\left<[({\rm Fe\,II,Cr\,II})/{\rm H}]\right>=-0.99\pm0.04\pm{\it 0.06}$, 
is consistent with the CMD stellar populations analysis.  

%
% agreement with nebular alpha abundances
%
The mean $\alpha$ element abundance from magnesium in the three stars is
$\left<[\alpha({\rm Mg\,I})/{\rm H}]\right>=-1.09\pm0.02\pm{\it 0.19}$
which is in good agreement with the nebular results for oxygen, 
$[\rm{O/H}] =-1.17$ \citep{Ski89} and $[\rm{O/H}] =-0.95$ \citep{Pil01}.
It is worth noting that this independent measurement of the $\alpha$
elemental abundance confirms the location of Sextans\,A in the 
fundamental metallicity -- luminosity relationship for dwarf 
galaxies \citep{Ski89, Ric95}.

The agreement between the $\alpha$ element abundances from A
supergiant stars and the nebular abundances is as expected. These
stars have ages $\approx10^7$ Myr and are expected to have formed out
of the same interstellar gas as seen in the nebulae.  Similar findings
are reported from the abundance analyses of B-type main sequence stars
in Orion and the LMC \citep{Cun94, Kor02}, respectively, the Galactic
abundance gradients from B stars \citep{Gum98, Rol00}, and A
supergiant analyses in the SMC, M31, and NGC\,6822 \citep{Ven99,
  Ven00, Ven01}, respectively.
It should be noted here, that the pristine nebular and stellar
abundances of some elements can be altered by effects of dust
depletion and rotational mixing which complicates a quantitative
comparison of nebular and stellar abundances \citep{Kor02}. However,
in the case of the stars and nebulae studied in Sextans\,A , the good
agreement within the estimated errors suggest that the contributions
due to these effects to the determined abundances are small.
Only the dIrr galaxy WLM shows a discrepancy between the stellar and
nebular $\alpha$ element abundances \citep{Ven03a}.   The nebulae
in WLM have $[{\rm O}/{\rm H}]=-0.89$ \citep{Ski89}, whereas two 
A supergiants have $[{\rm Mg}/{\rm H}]=-0.62$, and one star suggests 
$[{\rm O}/{\rm H}]=-0.21$.    These are significant differences that
are currently not understood and may suggest that WLM is the first
galaxy to display inhomogeneous mixing.

\begin{figure}
\includegraphics[angle=-90,width=\textwidth,clip]{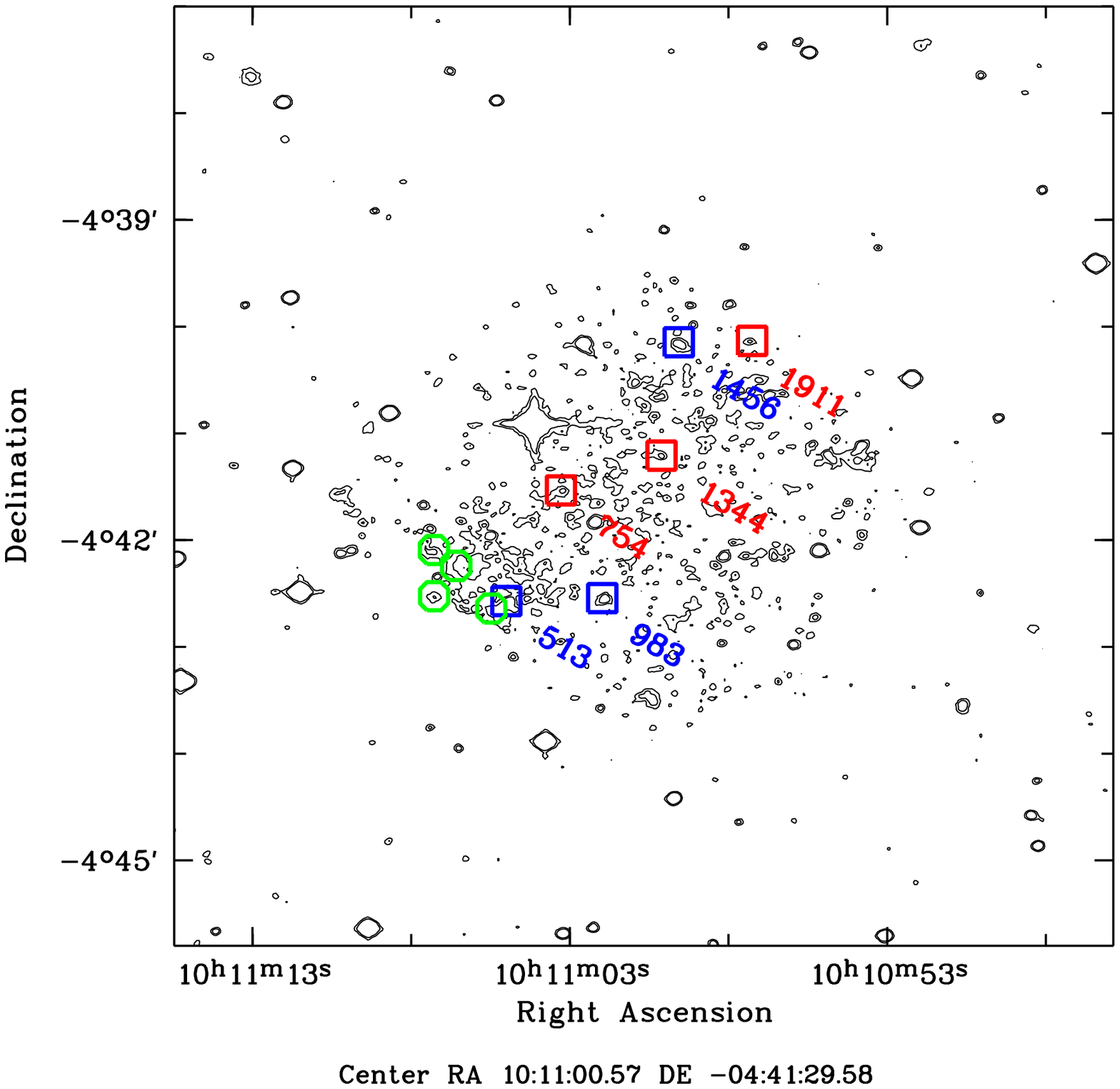}
\caption{\label{SexA-field} The Sextans A dwarf irregular galaxy.
  The three A supergiants analyzed in this work are indicated with
  red squares; other blue supergiants from the target list are 
  indicated with blue squares. The four bright compact \ion{H}{2} regions 
  analyzed in \citet{Ski89} are shown with green circles.}
\end{figure}

\begin{figure}
\includegraphics[angle=-90,width=\textwidth,clip]{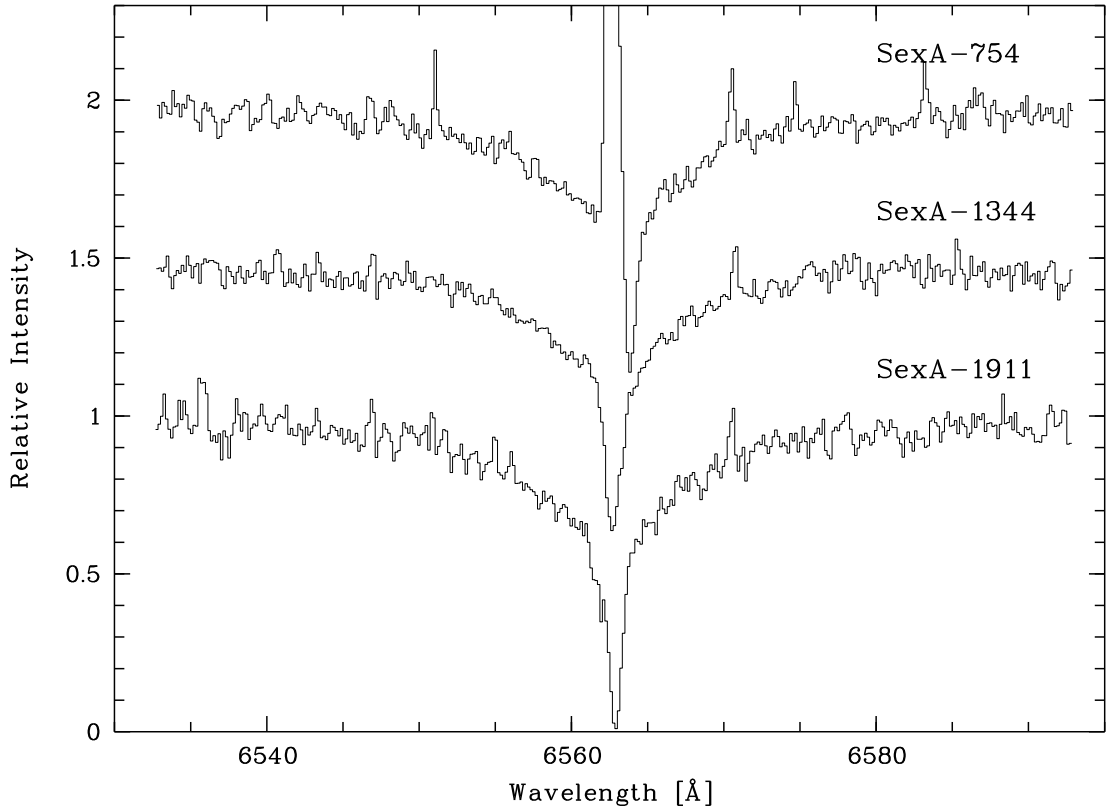}
\caption{\label{SexA-Halpha} H$\alpha$ profiles of the three
  Sextans\,A stars under analysis. All profiles have been shifted to
  the laboratory wavelength of H$\alpha$ by the respective
  heliocentric velocties from Tab.~\ref{SexA-sample}. The sharp,
  slightly blue-shifted emission seen in star SexA-754 is due to
  imperfect subtraction of the highly-structured diffuse H$\alpha$ 
  emission from the \ion{H}{1} gas of the galaxy.  Note,
  that no strong wind-emission features are present in the H$\alpha$
  profiles indicating the lower luminosity and/or metallicity of the
  stars.}
\end{figure}

\begin{figure}
\includegraphics[angle=-90,width=\textwidth,clip]{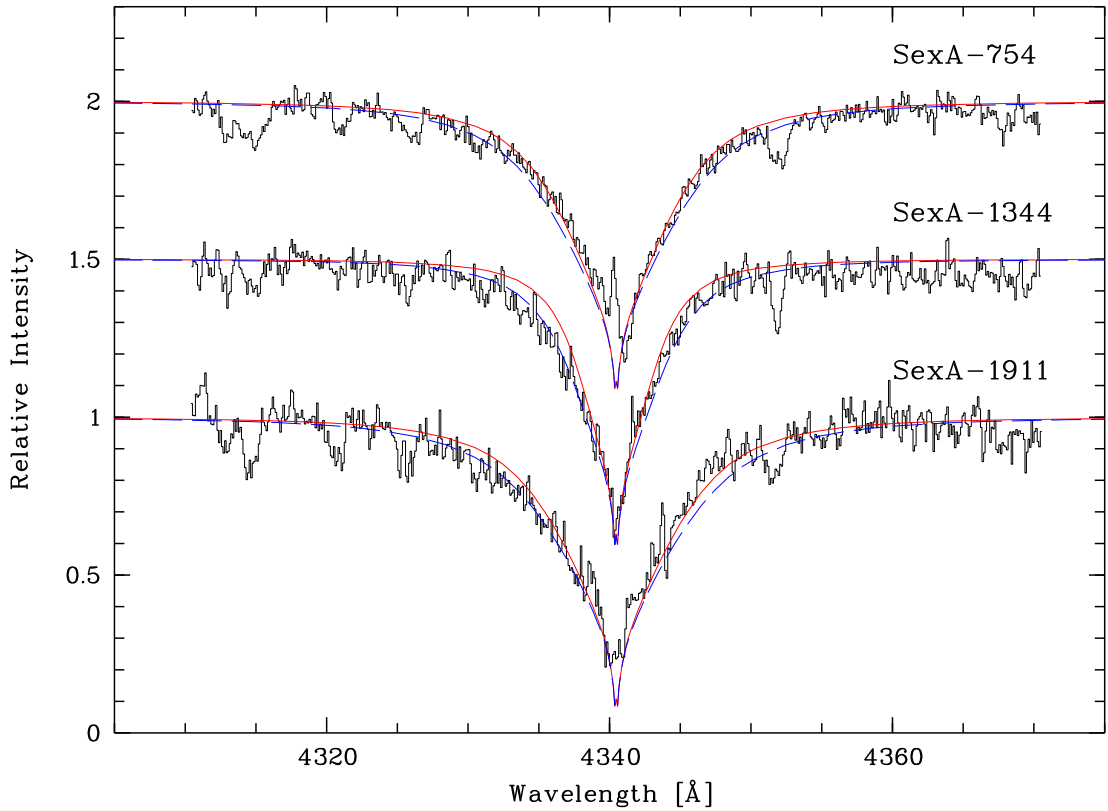}
\caption{\label{SexA-Hgamma} H$\gamma$ profiles of the three
  Sextans\,A stars under analysis. The solid (red) line indicates the
  fit of the synthetic profiles to the wings of the Balmer line for
  the final model atmosphere parameters while the dashed (blue) line
  is the synthetic profile for the same stellar parameters but with
  $\log g$ increased by $+0.2$\,dex which is considered the
  uncertainty in the fit.}
\end{figure}

\begin{figure}
\includegraphics[angle=-90,width=\textwidth,clip]{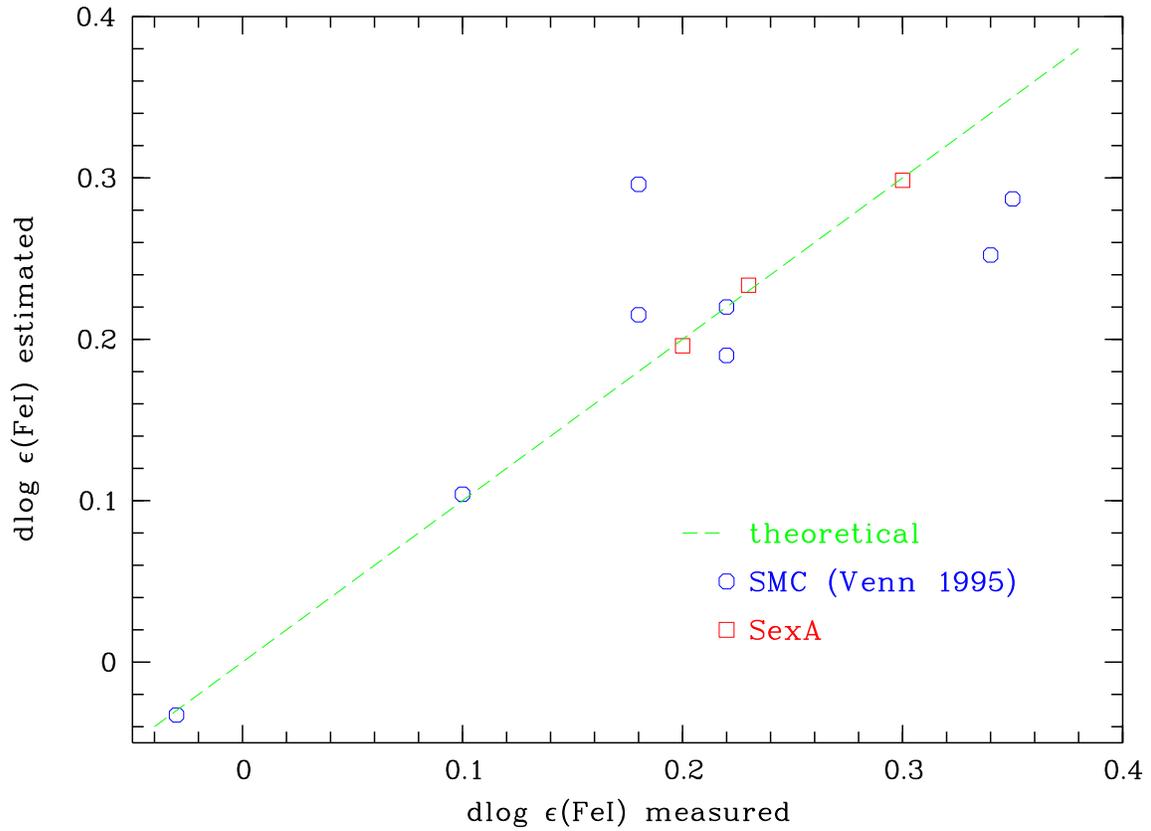}
\caption{\label{SexA-FeINLTE} The \ion{Fe}{1} NLTE corrections
  used in this work have been estimated using Eqn.~(\ref{eqnFeINLTE}).
  This relation was calibrated for the low metallicity and gravity of
  the stars in Sextans\,A (red squares) using the measured NLTE
  corrections in stars in the SMC (blue circles). The figure compares
  the measured corrections with the estimates from
  Eqn.~(\ref{eqnFeINLTE}).  }
\end{figure}

\begin{figure}
\includegraphics[angle=-90,width=\textwidth,clip]{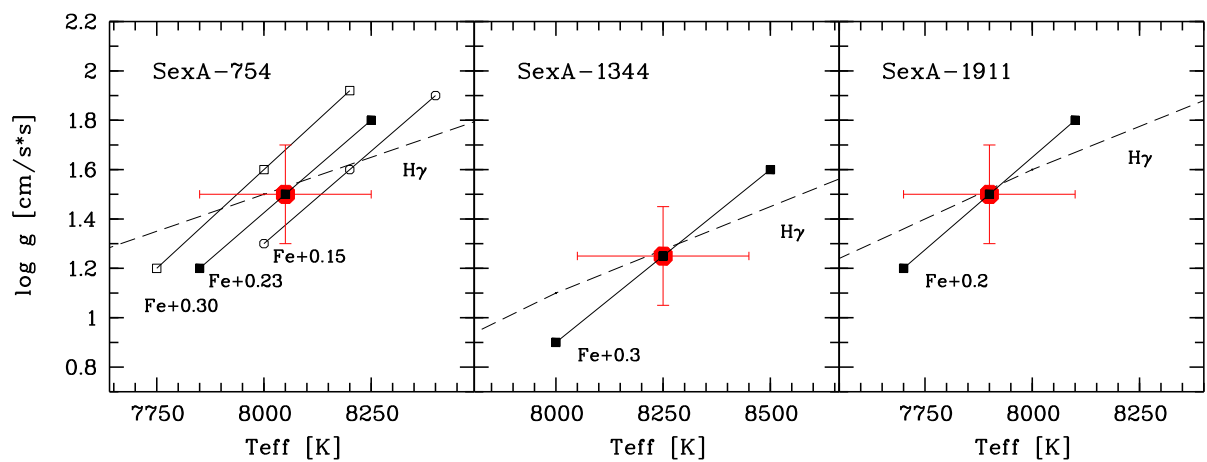}
\caption{\label{SexA-tefflogg} The final selected atmospheric parameters for
  the three analyzed A supergiants are indicated in the $T_{\rm eff} -
  \log g$ diagram by the (red) solid circles with errorbars and were
  determined by fits to the H$\gamma$ profiles (dashed line) and the
  \ion{Fe}{1}/\ion{Fe}{2} ionization equilibria (filled squares). In
  the labels ``Fe\ $x$'', the $x$ notes the NLTE correction applied to
  the \ion{Fe}{1} abundances. For star SexA-754 the ionization
  equilibria for three different \ion{Fe}{1} NLTE corrections are
  shown to illustrate the effect of the estimated uncertainties on
  the stellar parameters.}

\end{figure}

\begin{figure}
\includegraphics[angle=-90,width=\textwidth,clip]{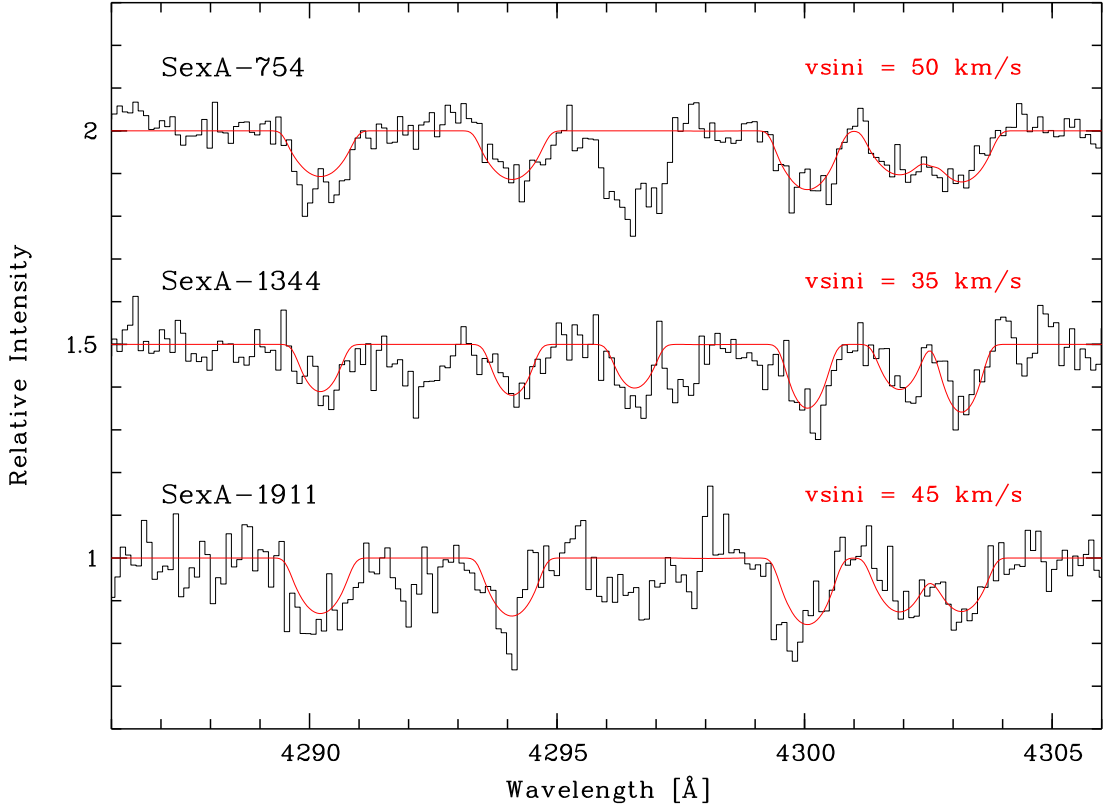}
\caption{\label{SexA-vsini} Spectrum synthesis to determine
  the projected rotational velocity $v\sin i$ of the three stars. The
  lines of \ion{Ti}{2}$\lambda\lambda4290,4294,4300,4301$ and the
  \ion{Fe}{2}$\lambda4303$ line where used to fit the profiles. Due
  to the lower $v\sin i$ of SexA-1344, the \ion{Fe}{2}$\lambda4297$
  line could be added to the list of lines to be fitted.  }
\end{figure}

\begin{figure}
\includegraphics[angle=-90,width=\textwidth,clip]{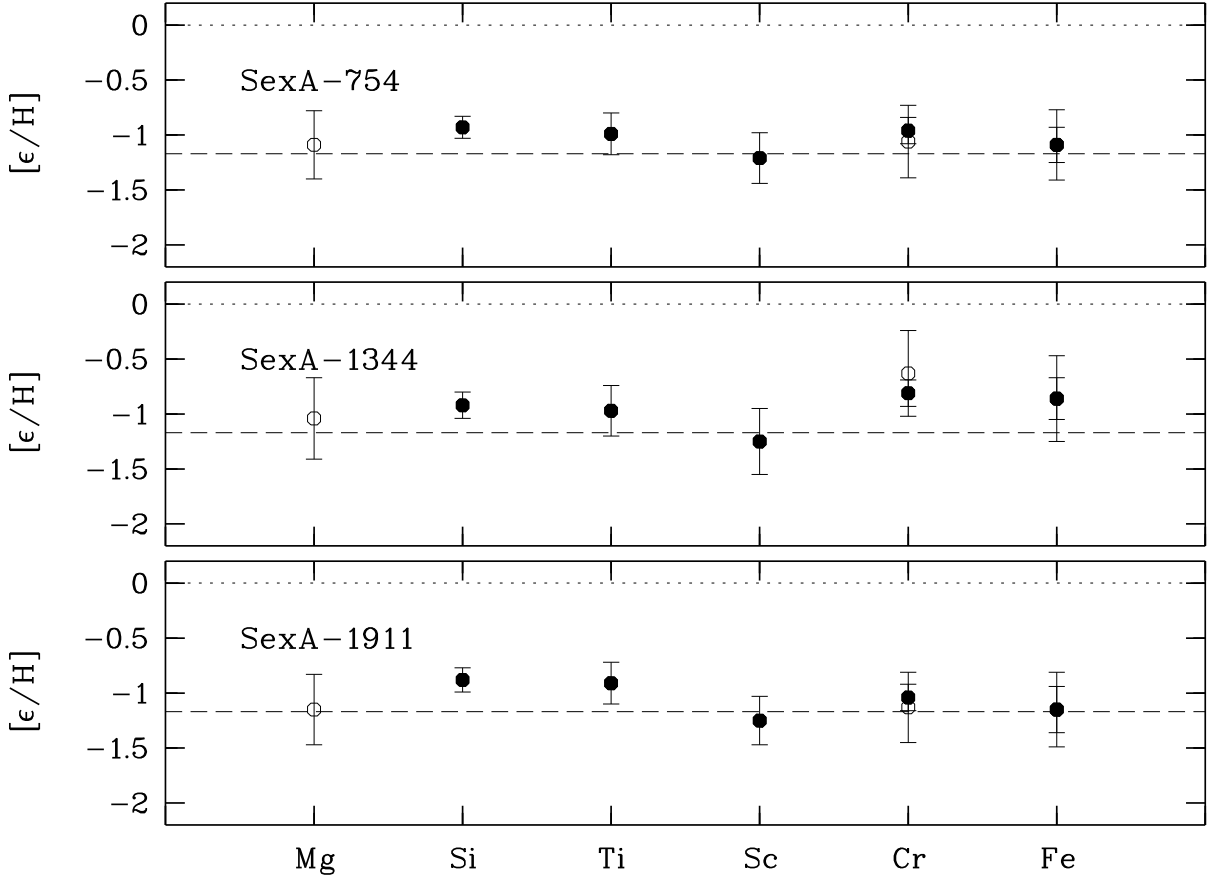}
\caption{\label{SexA-eps2solar} Elemental abundances for SexA-754 (top), 
  SexA-1344 (middle), and SexA-1911 (bottom) relative to solar (dotted
  line) and to the nebular oxygen underabundance of $-1.17$\,dex (dashed line).
  Abundances derived from neutral atoms are indicated with open circles, 
  abundances from singly ionized atoms with filled circles. The shown
  errorbars are the estimated combined errors from the uncertainties 
  of the determination of the stellar parameters 
  (cf.~Tab.~\ref{SexA-abuerror}).}
\end{figure}

\begin{figure}
\includegraphics[width=12cm,clip]{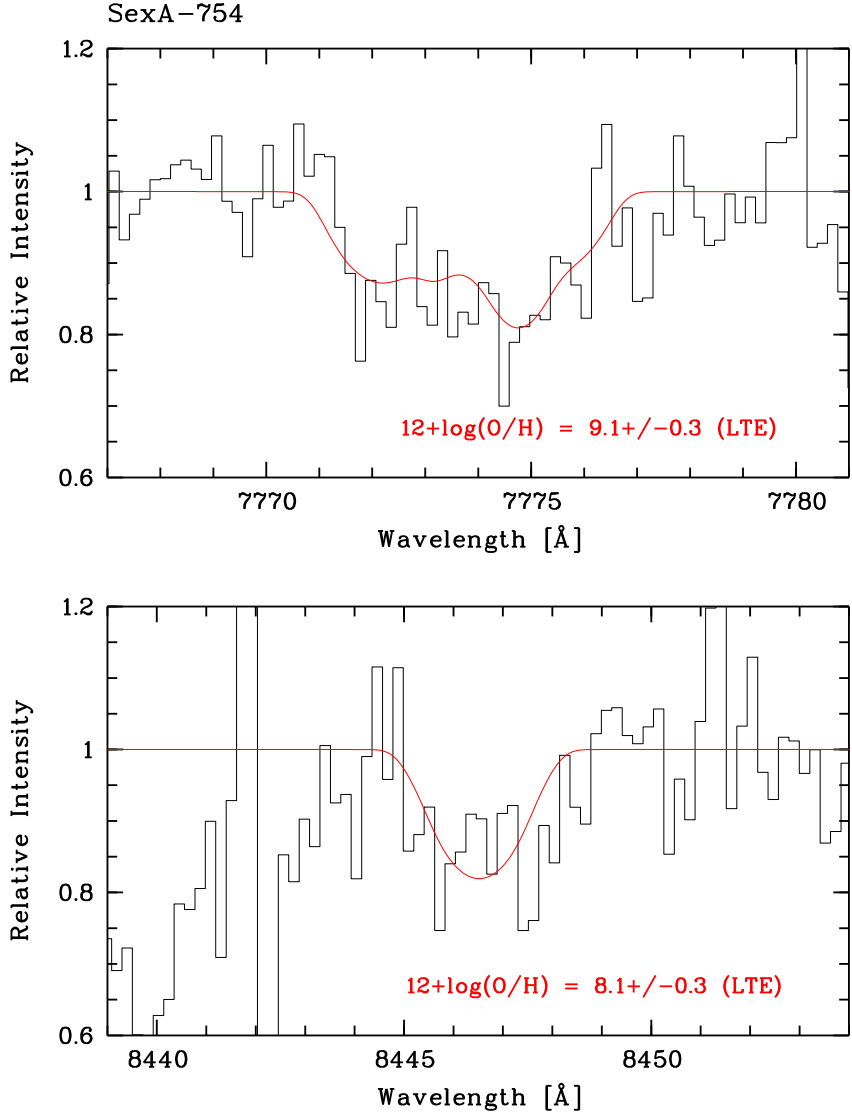}
\caption{\label{SexA-754_OI} Spectrum synthesis for the the near-infrared
  \ion{O}{1}$\lambda\lambda7771$-$5,8446$ multiplet lines. The oxygen abundance
  was fitted using spectrum synthesis in LTE. The uncertainty of the 
  LTE abundances of $\pm0.3$\,dex are due to the badly-defined continuum.
  To convert the measured LTE abundances into sensible oxygen abundances
  detailed computation of the expectedly large NLTE corrections \citep{Prz00}
  have to be carried out (see Section~\ref{abunds}).}
\end{figure}

%
% spatial distribution
%
The spatial distribution of abundances in a galaxy keeps an important
record of the re-distribution history of freshly produced elements 
and their mixing with the ISM.
As can be seen from Fig.~\ref{SexA-field}, the three stars analyzed in
this work and the four bright compact \ion{H}{2} regions examined by
\citet{Ski89} lie on a line across the central, visually
brightest part of the galaxy.   Adopting D$_{\rm Sex\,A} = 1.3$\,Mpc
\citep{Dol03a}, no significant $\alpha$ element abundance variations 
are found over a distance of $0.8$\,kpc as covered by the stars 
(between SexA-754 to SexA-1344 to SexA-1911), 
and over $1.6$\,kpc if the \ion{H}{2} regions are included. 
The same level of abundance homogeneity is found
for the iron-group elements in the stars.
%
%However, possibly a small enhancement of some $0.25$\,dex in the
%iron-group elements of SexA-1344, i.e., the star closest to the
%center of the galaxy (TBC) was found. However, this deviation is
%within $2\sigma$ of the estimated systematic errors and therefore
%most likely not significant.
%
The high level of homogeneity of nebular abundances in dwarf irregular
galaxies has been discussed as possible evidence against in situ (``on
the spot'') enrichment by \citet{Kob97} and therefore against the
general validity of the instantaneous recycling approximation as
typically used in chemical evolution models. \citet{Kob97} favour a
scenario in which the newly synthesized elements remain in either a
difficult to observe hot $10^6$\,K or an equally difficult to observe
cold, dusty phase while mixing throughout the galaxy.  Recent Chandra
spectroscopy results for NGC\,1569 by \citet{Mar02} favor the hot
phase as the reservoir for newly synthesized and expelled gas.

%
% alpha/Fe 
%
The $\alpha/{\rm Fe}$ abundance ratio is a key constraint for the
chemical evolution of a galaxy. $\alpha$ elements are primarily
synthesized in SNe\,II while the iron-group elements are mainly
produced in SNe\,Ia but also in SNe\,II.
A star formation burst in a galaxy will temporarily increase the
$\alpha/{\rm Fe}$ ratio of the ISM due to the massive stars that
quickly enrich the ISM with $\alpha$ elements, while the lower mass
stars will then slowly increase the content of iron-group elements
resulting in a slow decline of the $\alpha/{\rm Fe}$ ratio of the
galaxy \citep{Tin79}.  
\citet{Gil91} demonstrated the expected differences 
in the $\alpha/{\rm Fe}$ ratios
for galaxies with different star formation histories, and in
particular noted that the solar $\alpha/{\rm Fe}$ ratio must not be
regarded as an universal ratio.

From the three stars in Sextans\,A, 
the $\alpha/{\rm Fe}$ ratio is measured as 
$\left<[\alpha({\rm Mg\,I})/{\rm
    Fe\,II,Cr\,II}]\right>=-0.11\pm0.02\pm{\it 0.10}$.
This $\alpha$/Fe ratio is much lower than in Galactic stars
of similar metallicity \citep{Edv93,Nis97}.   At the metallicity
of these Sextans\,A stars, [Fe/H] $\approx -1$,
the Galactic thick disk and Galactic halo stars overlap 
with a [$\alpha$/Fe] $\approx$ +0.3.
The lower $\alpha$/Fe ratios in these Sextans\,A stars further follows
the trend found from the abundance analyses of stars in two other
dIrrs, NGC\,6822 and WLM \citep{Ven01, Ven03a}, but extends the
metallicity range now sampled from $[{\rm Fe}/{\rm H}]=-0.5$ in those
galaxies to $[{\rm Fe}/{\rm H}]=-1.0$.  As discussed by
\citet{Ven03b}, this lower metallicity now overlaps the upper
metallicities sampled by red giant stars in dwarf spheroidal galaxies,
which also show much lower $\alpha$/Fe ratios than Galactic metal-poor
stars \citep{She01,She03,Tol03}.  In the dwarf spheroidal galaxies,
the low [$\alpha$/Fe] ratios near [Fe/H] = $-1.0$ are most likely due
to the lower star formation rates, thus slower chemical evolution.
The metallicity range of the dwarf irregular galaxies also overlaps
with the damped Ly$\alpha$ absorption systems \citep{Nis03, Led02,
  Pet99}, which are also recognized to have low $\alpha$/Fe ratios.
This might suggest that low $\alpha$/Fe ratios are a generic effect 
at low metallicities.

The star formation histories for Sextans\,A, WLM, and NGC\,6822 are 
globally similar as derived from their CMDs.   A comparison can be
seen in Fig.~8 of \citet{Mat98}. Accordingly, all
three dIrrs had significant star formation at ancient times, 
$>5-10$\,Gyr ago, with a hiatus at intermediate ages, $1-5$\,Gyr ago, 
and recent star formation events in the past 1\,Gyr.  
\citet{Mat03} has recently emphasized that the SFH of a galaxy affects
the {\it absolute} elemental abundances, but is only of minor importance 
for the abundance {\it ratios}, the latter being primarily
determined by the stellar lifetimes, initial mass functions (IMF) and
the stellar nucleosynthesis.   Thus, 
the impact of different SFHs on the $\alpha/{\rm Fe}$ ratio is 
related to the star formation efficiency, and manifests itself as
a short plateau in the $[\alpha/{\rm Fe}]$ ratio at low metallicities 
in galaxies with low star formation rates, like irregulars 
% spirals?  which parts since we then compare the dIrr to the
% Galactic metal-poor stars and say they are different
(either in bursts or continuous), while the plateau is maintained 
to higher metallicities for systems with high star formation rates 
like bulges and ellipticals.
Thus, our finding that the [$\alpha/{\rm Fe}$] ratios in three dIrrs 
with similar SFHs is significantly lower than in the metal-poor 
Galactic stars can be explained in this scenario.
The three dIrrs with a metallicity range from 
$-1<[{\rm Fe}/{\rm H}]<-0.4$ have already left
the $[\alpha/{\rm Fe}]$ plateau and presently display the same lower
$[\alpha/{\rm Fe}]$ ratio.   That this ratio is similar to solar
suggests that the integrated stellar yields from a star formation
event is consistent with the solar abundance ratios, probably due 
to the {\it universal} nature of stellar lifetimes, IMFs, and 
stellar nucleosynthesis.      
That the high-redshift DLA absortion systems also show low
[$\alpha$/Fe] ratios suggests that they are also systems with
low star formation rates, but this could be either irregular
galaxies or the outer parts of spirals.
A recent study of published DLA
system abundances in comparison with chemical evolution models of
different galaxy morphologies by \citet{Cal03} has identified the
irregular and spiral galaxies as the possibly ideal sites to create
the DLA systems, while ellipticals can be ruled out.

% what about variable IMFs, or gas outflows at various epochs?

The value of high resolution spectroscopy of the brightest blue stars 
in nearby galaxies for detailed, quantitative abundance studies has been 
further demonstrated in this work.   To gain further insight into the 
chemical evolution of the dwarf galaxies and hence into the early 
evolution of galaxies with low star formation rates, requires more 
detailed abundance analyses of their stars, particularly towards 
the lower metallicities.

%
%%%%%%%%%%%%%%%%%%%%%%%%%%%%%%%%%%%%%%%%%%%%%%%%%%%%%%%%%%%%%%%%%%%%%%%%%%%
%
\acknowledgments

We thank H.A.~Kobulnicky for the low resolution Keck spectrum of
SexA-754 which allowed us to get this work started. AK wants to thank
the ESO Director General's Discretionary Fund (DGDF) for the support
of this project and the Institute of Astronomy, University of
Cambridge, UK for its hospitality during a short-term visit where a
major part of this work was done. KAV would like to thank the NSF for
support through a CAREER award, AST-9984073, and the IoA Cambridge for
support during a one year visit. CP wants to thank ESO for the
possibility of a short-term studentship at the ESO facilities in
Vitacura, Santiago de Chile. ET gratefully acknowledges support from a
fellowship of the Royal Netherlands Academy of Arts and Sciences.
Further we want to thank the referee A.J.Korn for valuable comments
on the manuscript.
%
%%%%%%%%%%%%%%%%%%%%%%%%%%%%%%%%%%%%%%%%%%%%%%%%%%%%%%%%%%%%%%%%%%%%%%%%%%%
%
%\appendix
%\section{Appendicial material}

%
%%%%%%%%%%%%%%%%%%%%%%%%%%%%%%%%%%%%%%%%%%%%%%%%%%%%%%%%%%%%%%%%%%%%%%%%%%%
%

%
%%%%%%%%%%%%%%%%%%%%%%%%%%%%%%%%%%%%%%%%%%%%%%%%%%%%%%%%%%%%%%%%%%%%%%%%%%%
%
\clearpage


\clearpage 

\clearpage 

\clearpage 

\clearpage

\clearpage 

\clearpage 

\clearpage 

%
%%%%%%%%%%%%%%%%%%%%%%%%%%%%%%%%%%%%%%%%%%%%%%%%%%%%%%%%%%%%%%%%%%%%%%%%%%%
%
\clearpage 

\begin{deluxetable}{lcccccc}
\tabletypesize{\small}
\tablecaption{\label{SexA-observations}Sextans\,A Keck/ESI and 
  VLT/UVES observations}
\tablewidth{0pt}
\tablecolumns{7}
\tablehead{
\colhead{Star}                 & 
\colhead{Date}                 & 
\colhead{UTC}                  &
\colhead{$\lambda_{\rm c}$}    &
\colhead{Exp.Time}             &
\colhead{Airmass}              & 
\colhead{Seeing}       \\
\colhead{ }                    & 
\colhead{ }                    & 
\colhead{Start}                &
\colhead{[nm]}                 &
\colhead{[sec]}                &
\colhead{Start}                & 
\colhead{[$\arcsec$]}       
}
\startdata
\sidehead{Keck/ESI} 
SexA-513 
   &    2002-02-13 &  08:45 &  750     & 2700 & 1.30 & 1.5   \\
   &    2002-02-14 &  07:49 &  750     & 2700 & 1.38 & 1.6   \\
SexA-983 
   &    2002-02-13 &  09:44 &  750     & 3600 & 1.15 & 1.5   \\
SexA-1344
   &    2002-02-14 &  09:59 &  750     & 3600 & 1.10 & 1.6   \\
SexA-1456
   &    2002-02-13 &  11:02 &  750     & 2600 & 1.10 & 1.5   \\
\sidehead{VLT/UVES} 
SexA-754 
   &    2002-01-11 &  04:45 &  390+580 & 4500 & 1.39 & 0.98  \\
   &    2002-01-13\tablenotemark{a} 
                   &  05:55 &  390+580 & 4500 & 1.14 & 1.50  \\
   &    2002-01-13 &  07:12 &  390+580 & 4500 & 1.06 & 1.20  \\
   &    2002-01-18 &  04:55 &  390+580 & 4500 & 1.24 & 1.25  \\
   &    2002-01-20 &  07:17 &  390+580 & 4500 & 1.06 & 0.97  \\
   &    2002-01-23 &  05:55 &  390+580 & 4500 & 1.08 & 0.81  \\
   &    2002-02-08 &  02:00 &  390+580 & 4500 & 1.80 & 1.19  \\
   &    2002-02-09 &  02:32 &  390+580 & 4500 & 1.50 & 0.73  \\
   &    2002-02-10 &  01:55 &  390+580 & 4500 & 1.77 & 0.75  \\
   &    2002-02-11 &  02:41 &  390+580 & 4500 & 1.40 & 0.58  \\
   &    2002-02-12 &  02:14 &  390+580 & 4500 & 1.55 & 1.13  \\[0.5ex]
   &    2002-04-14 &  00:58 &  437+840 & 4500 & 1.07 & 1.00  \\
   &    2002-04-14 &  02:15 &  437+840 & 4500 & 1.08 & 1.02  \\
   &    2002-04-15 &  00:30 &  437+840 & 4500 & 1.08 & 0.80  \\
   &    2002-04-15 &  01:48 &  437+840 & 3600 & 1.07 & 0.67  \\
   &    2002-04-15 &  02:50 &  437+840 & 3600 & 1.14 & 0.87  \\
   &    2002-04-15 &  03:51 &  437+840 & 3600 & 1.33 & 1.01  \\[1ex]
SexA-1344 
   &    2002-12-30 &  05:19 &  390+580 & 4500 & 1.48 & 1.25  \\
   &    2002-12-30 &  06:36 &  390+580 & 4500 & 1.17 & 1.03  \\
   &    2003-01-01 &  07:24 &  390+580 & 4500 & 1.08 & 1.42  \\
   &    2003-01-08 &  04:39 &  390+580 & 4500 & 1.50 & 0.84  \\
   &    2003-01-08 &  05:58 &  390+580 & 4500 & 1.18 & 1.08  \\
   &    2003-01-08 &  07:16 &  390+580 & 4500 & 1.07 & 0.61  \\
   &    2003-02-04 &  03:12 &  390+580 & 4500 & 1.39 & 1.00  \\
   &    2003-02-04 &  04:29 &  390+580 & 4500 & 1.14 & 1.27  \\
   &    2003-02-04 &  05:46 &  390+580 & 4500 & 1.06 & 1.15  \\
   &    2003-02-05 &  04:38 &  390+580 & 4500 & 1.12 & 1.29  \\[1ex]
SexA-1911 
   &    2003-01-12 &  04:58 &  390+580 & 4500 & 1.32 & 1.19  \\
   &    2003-02-05 &  05:59 &  390+580 & 4500 & 1.06 & 1.02  \\
   &    2003-02-05 &  07:19 &  390+580 & 4500 & 1.14 & 1.17  \\
   &    2003-02-06 &  03:43 &  390+580 & 4500 & 1.23 & 0.93  \\
   &    2003-02-07 &  04:13 &  390+580 & 4500 & 1.15 & 0.78  \\
   &    2003-02-07 &  05:29 &  390+580 & 4500 & 1.06 & 0.95  \\
   &    2003-02-07 &  07:49 &  390+580 & 4500 & 1.23 & 0.93  \\
   &    2003-02-08 &  04:07 &  390+580 & 4500 & 1.15 & 1.22  \\
   &    2003-02-08 &  06:22 &  390+580 & 4500 & 1.08 & 1.09  \\
   &    2003-02-08 &  07:40 &  390+580 & 4500 & 1.21 & 0.94  \\[1ex]
SexA-513 
   &    2002-04-13 &  23:19 &  390+580 & 1800 & 1.23 & 1.12  \\
   &    2002-04-14 &  23:23 &  390+580 & 1800 & 1.21 & 0.91  \\
   &    2002-04-14 &  00:12 &  390+580 & 1800 & 1.14 & 0.72  \\[1ex]
SexA-983 
   &    2002-04-13 &  23:53 &  390+580 & 3600 & 1.15 & 1.13  \\
   \enddata \tablenotetext{a}{Observation was repeated in the next
     exposure because the seeing constraint was violated here}
   \tablecomments{ All ESI data was taken in echelle mode with a
     $20\arcsec \times 1.25\,\arcsec$ slit corresponding to a
     resolving power of $R=3200$.  The wavelength coverage is
     $390-1100$\,nm.  All UVES data was taken in dichroic mode, with a
     $1.2\,\arcsec$ slit width and 2x2 CCD binning corresponding to a
     3-pix resolving power of $R=40\,000$. The wavelength coverage of
     the dichroic settings is for 390+580: $330-452,\ 478-575,\ 
     584-680$\,nm and for 437+840: $376-498,\ 650-831,\ 845-1025$\,nm.
     }
\end{deluxetable}

%\clearpage 

\begin{deluxetable}{lcccc}
\tabletypesize{\small}
\tablecaption{\label{SexA-snr}Signal-to-Noise Ratios of 
  the combined UVES spectra}
\tablewidth{0pt}
\tablehead{
\colhead{Star} & 
\multicolumn{4}{c}{Signal-to-Noise Ratio\tablenotemark{a}} \\
\colhead{    } & 
\colhead{@400\,nm} & 
\colhead{@500\,nm} & 
\colhead{@650\,nm} & 
\colhead{@800\,nm} 
}
\startdata
SexA-754   &    35   &    45   &    35   &  10       \\
SexA-1344  &    30   &    35   &    30   &  \nodata  \\
SexA-1911  &    25   &    35   &    25   &  \nodata  \\
SexA-513   &    30   &    35   &    30   &  \nodata  \\
SexA-983   &    15   &    20   &    15   &  \nodata  \\
\enddata
\tablenotetext{a}{For 2\,pix resolution elements of $R=20\,000$}
\end{deluxetable}

\clearpage 

\begin{deluxetable}{lrrcll}
\tabletypesize{\small}
\tablecaption{\label{SexA-sample}Sextans\,A sample}
\tablewidth{0pt}
\tablecolumns{6}
\tablehead{
\colhead{Star\tablenotemark{a}} & 
\colhead{$V$\tablenotemark{a}} & 
\colhead{$(B-V)$\tablenotemark{a}}  &
\colhead{RV$_{\rm hel}$[km/s]} &
\colhead{Sp.Type} &
\colhead{Comment}
}
\startdata
\sidehead{analyzed targets} 
SexA-754  & $19.5$ & $-0.014$ & $+340$  &  A7\,Iab supergiant\tablenotemark{b}  & UVES (cand. 2)     \\
SexA-1344 & $19.4$ & $-0.083$ & $+320$  &  A7\,Iab supergiant\tablenotemark{b}  & ESI+UVES (cand. 4) \\
SexA-1911 & $19.6$ & $-0.004$ & $+320$  &  A8\,Iab supergiant\tablenotemark{b}  & UVES (cand. 5)     \\
\sidehead{discarded targets} 
SexA-513  & $17.5$ & $-0.031$ & $+350$  &  F hypergiant        & ESI+UVES (cand. 1) \\
SexA-983  & $18.3$ & $-0.011$ & $+290$  &  G supergiant        & ESI+UVES (cand. 3) \\
SexA-1456 & $18.8$ & $+0.254$ & $+320$  &  K supergiant        & ESI (cand. 6)      \\
\enddata
\tablenotetext{a}{From \citet{vDyk98}}
\tablenotetext{b}{As derived in this work, cf. Tab.~\ref{SexA-analysis}}
\end{deluxetable}

\clearpage 

\begin{deluxetable}{lcccc}
\tabletypesize{\small}
\tablecaption{\label{SexA-analysis}Atmospheric Analysis}
\tablewidth{0pt}
\tablecolumns{5}
\tablehead{
\colhead{ }                    & 
\colhead{Solar}                &
\colhead{SexA-754}             & 
\colhead{SexA-1344}            & 
\colhead{SexA-1911}            
}
\startdata
%                         & &  SexA-754    &  SexA-1344   & SexA-1911      \\
$T_{\rm eff}$\,[K]        & & $8050\pm200$ & $8250\pm200$ & $7900\pm200$   \\
$\log g$\,[cm/s$^2$]      & & $1.5\pm0.2$  & $1.25\pm0.2$ & $1.5\pm0.2$    \\
$\xi_{\rm micro}$\,[km/s] & & $4.0\pm1$    & $3.5\pm1$    & $3.5\pm1$      \\
$v\sin i$\,[km/s]         & & $50\pm5$     & $35\pm5$     & $45\pm5$       \\
$\log(L/{\rm L}_\odot)$   & & $4.35\pm0.2$ & $4.39\pm0.2$ & $4.31\pm0.2$   \\
$R/{\rm R}_\odot$         & & $77\pm20$    & $77\pm20$    & $76\pm20$      \\
Sp.Type                   & &  A7\,Iab     &  A7\,Iab     &  A8\,Iab       \\[1.5ex]
Mg\,I (NLTE) & 
$7.58$ & 
$6.49 \pm 0.01\ (4) \pm {\it 0.30}$ & 
$6.54 \pm 0.04\ (4) \pm {\it 0.34}$ & 
$6.43 \pm 0.05\ (4) \pm {\it 0.31}$ \\
%{\it Mg\,II NLTE}\tablenotemark{a} & 
%${\it 7.58}$ & 
%${\it 6.64} \pm {\it 0.17 \ (5)} \pm {\it 0.07}$ & 
%${\it 6.54} \pm {\it 0.00 \ (1)} \pm {\it 0.30}$ & 
%${\it 6.47} \pm {\it 0.00 \ (1)} \pm {\it 0.26}$ \\
Mg\,II (NLTE) & 
$7.58$ & 
$<6.62 \pm 0.18\ (4) \pm {\it 0.07}$ & 
\nodata & 
\nodata \\
Si\,II (LTE)   & 
$7.56$ & 
$6.63 \pm 0.13\ (9) \pm {\it 0.10}$ & 
$6.64 \pm 0.08\ (9) \pm {\it 0.12}$ & 
$6.68 \pm 0.09\ (9) \pm {\it 0.11}$ \\[1ex]
Ti\,II (LTE)   & 
$4.94$ & 
$3.95 \pm 0.16 (37) \pm {\it 0.19}$ & 
$3.97 \pm 0.12 (30) \pm {\it 0.23}$ & 
$4.03 \pm 0.17 (30) \pm {\it 0.19}$ \\
Sc\,II (LTE)   & 
$3.10$ & 
$1.89 \pm 0.06\ (6) \pm {\it 0.23}$ & 
$1.85 \pm 0.07\ (4) \pm {\it 0.29}$ & 
$1.85 \pm 0.05\ (5) \pm {\it 0.22}$ \\[1ex]
Cr\,I (LTE)    & 
$5.69$ & 
$4.63 \pm 0.14\ (3) \pm {\it 0.31}$ & 
$5.06 \pm 0.00\ (1) \pm {\it 0.37}$ & 
$4.56 \pm 0.00\ (1) \pm {\it 0.31}$ \\
Cr\,II (LTE)   & 
$5.69$ & 
$4.73 \pm 0.13 (20) \pm {\it 0.12}$ & 
$4.88 \pm 0.07\ (8) \pm {\it 0.12}$ & 
$4.65 \pm 0.12 (13) \pm {\it 0.11}$ \\
Fe\,I (LTE)    & 
$7.50$ & 
$6.18 \pm 0.13 (15) \pm {\it 0.31}$ & 
$6.34 \pm 0.06\ (7) \pm {\it 0.37}$ & 
$6.15 \pm 0.04 (12) \pm {\it 0.33}$ \\
Fe\,II (LTE)   & 
$7.50$ & 
$6.41 \pm 0.12 (14) \pm {\it 0.16}$ & 
$6.64 \pm 0.05 (15) \pm {\it 0.19}$ & 
$6.35 \pm 0.06 (12) \pm {\it 0.21}$ \\
\enddata
%
%\tablenotetext{a}{Not used in any of the analyses (cf. text)}
\end{deluxetable}

\clearpage 

\begin{deluxetable}{lrrrrrrrrr}
\tabletypesize{\small}
\tablecaption{\label{SexA-abuerror}Abundance Uncertainties}
\tablewidth{0pt}
\tablecolumns{10}
\tablehead{
\colhead{ }                    & 
\multicolumn{3}{c}{SexA-754}   & 
\multicolumn{3}{c}{SexA-1344}  & 
\multicolumn{3}{c}{SexA-1911}  \\[1ex]
\colhead{ }                    & 
\colhead{$\Delta T_{\rm eff}$} & 
\colhead{$\Delta \log g$}      & 
\colhead{$\Delta \xi$}         & 
\colhead{$\Delta T_{\rm eff}$} & 
\colhead{$\Delta \log g$}      & 
\colhead{$\Delta \xi$}         & 
\colhead{$\Delta T_{\rm eff}$} & 
\colhead{$\Delta \log g$}      & 
\colhead{$\Delta \xi$}         \\
\colhead{ }                    & 
\colhead{$+200$\,K}            & 
\colhead{$+0.2$}               & 
\colhead{$+1$\,km/s}           & 
\colhead{$+200$\,K}            & 
\colhead{$+0.2$}               & 
\colhead{$+1$\,km/s}           & 
\colhead{$+200$\,K}            & 
\colhead{$+0.2$}               & 
\colhead{$+1$\,km/s}           \\
\colhead{$\Delta\log(X/{\rm H})$} & \multicolumn{9}{c}{\ }           
}
\startdata
%      & SexA-754               &  SexA-1344              &  SexA-1911             \\ 
Mg I   &$+0.28$&$-0.10$&$-0.10$ & $+0.33$&$-0.16$&$-0.05$ & $+0.27$&$-0.10$&$-0.15$\\
Mg II  &$+0.01$&$+0.02$&$-0.07$ & $+0.06$&$-0.04$&$-0.29$ & $+0.03$&$+0.00$&$-0.26$\\
Si II  &$-0.01$&$+0.02$&$-0.10$ & $+0.00$&$+0.02$&$-0.12$ & $+0.01$&$+0.02$&$-0.11$\\
Ti II  &$+0.18$&$+0.02$&$-0.06$ & $+0.21$&$-0.04$&$-0.08$ & $+0.17$&$+0.02$&$-0.09$\\
Sc II  &$+0.22$&$+0.00$&$-0.05$ & $+0.28$&$-0.08$&$-0.05$ & $+0.20$&$+0.00$&$-0.09$\\ 
Cr I   &$+0.31$&$-0.10$&$-0.01$ & $+0.36$&$+0.16$&$-0.02$ & $+0.31$&$-0.08$&$-0.03$\\
Cr II  &$+0.11$&$+0.02$&$-0.04$ & $+0.11$&$-0.02$&$-0.04$ & $+0.11$&$+0.02$&$-0.03$\\
Fe I   &$+0.30$&$-0.10$&$-0.06$ & $+0.35$&$+0.16$&$-0.08$ & $+0.30$&$-0.08$&$-0.13$\\
Fe II  &$+0.12$&$+0.02$&$-0.11$ & $+0.12$&$-0.02$&$-0.15$ & $+0.13$&$+0.02$&$-0.16$\\
\enddata
%\tablenotetext{a}{ }}
\end{deluxetable}

\clearpage 

\begin{deluxetable}{ccccc
cc
cc
cc
}
\tabletypesize{\small}
\tablecaption{\label{lines}Line Strengths, Atomic Data and Abundances}
\tablewidth{0pt}
\tablecolumns{11}
\tablehead{
\colhead{ }           &
\colhead{$\lambda$}   &
\colhead{Multiplet}   &
\colhead{$\chi$}      &
& \multicolumn{2}{l}{\hspace{1mm} SexA-754}
& \multicolumn{2}{l}{\hspace{1mm} SexA-1344}
& \multicolumn{2}{l}{\hspace{1mm} SexA-1911}
\\
\colhead{Element}     &
\colhead{(\AA)}       &
\colhead{Table}       &
\colhead{(eV)}        &
\colhead{$\log(gf)${\hspace{10 mm}}}
& \colhead{$W_{\lambda}$}
& \colhead{$\log(\epsilon)$}
& \colhead{$W_{\lambda}$}
& \colhead{$\log(\epsilon)$}
& \colhead{$W_{\lambda}$}
& \colhead{$\log(\epsilon)$}
}
\startdata
\sidehead{O I LTE line profile synthesis}
$800$ & $7772.20$ & $1$ & $9.15$ & $0.37$ & & & \\[-0.8ex]
$800$ & $7774.17$ & $1$ & $9.15$ & $0.22$ & {\it 700} & $ 9.1$  
& \nodata& \nodata  
& \nodata& \nodata  \\[-0.8ex]
$800$ & $7775.39$ & $1$ & $9.15$ & $0.00$ & & & 
\\
$800$ & $8846.25$ & $4$ & $9.52$ & $-0.46$ & & & \\[-0.8ex]
$800$ & $8846.36$ & $4$ & $9.52$ & $0.24$& {\it 385} & $ 8.1$  
& \nodata& \nodata  
& \nodata& \nodata  \\[-0.8ex]
$800$ & $8846.76$ & $4$ & $9.52$ & $0.01$ & & & 
\\
\sidehead{Mg I NLTE}
$1200$ & $3829.360$ & $3$ & $2.71$ & $-0.21$ 
& $ 140$ & $ 6.50$  
& $  80$ & $ 6.49$  
& $ 140$ & $ 6.42$  
\\
$1200$ & $5167.320$ & $2$ & $2.71$ & $-0.86$ 
& $  75$ & $ 6.49$  
& $  35$ & $ 6.54$  
& $  85$ & $ 6.40$  
\\
$1200$ & $5172.680$ & $2$ & $2.71$ & $-0.38$ 
& $ 135$ & $ 6.50$  
& $  80$ & $ 6.58$  
& $ 150$ & $ 6.50$  
\\
$1200$ & $5183.600$ & $2$ & $2.72$ & $-0.16$ 
& $ 160$ & $ 6.49$  
& $ 100$ & $ 6.55$  
& $ 160$ & $ 6.38$  
\\
\sidehead{Mg II NLTE}
$1201$ & $3848.210$ & $5$ & $8.86$ & $-1.56$ 
& $ <20$ & $<6.84$  
& $ $\nodata$ $ & $ $\nodata$ $  
& $ $\nodata$ $ & $ $\nodata$ $  
\\
$1201$ & $4390.570$ & $10$ & $10.0$ & $-0.50$ 
& $ <20$ & $<6.52$  
& $ $\nodata$ $ & $ $\nodata$ $  
& $ $\nodata$ $ & $ $\nodata$ $  
\\
%$1201$ & $4481.230$ & $4$ & $8.86$ & $0.97$ 
%& $ 240$ & $ 6.73$  
%& $ 200$ & $ 6.54$  
%& $ 200$ & $ 6.47$  
%\\
$1201$ & $7877.050$ & $8$ & $10.0$ & $0.39$ 
& $ <60$ & $<6.69$  
& $ $\nodata$ $ & $ $\nodata$ $  
& $ $\nodata$ $ & $ $\nodata$ $  
\\
$1201$ & $7896.370$ & $8$ & $10.0$ & $0.65$ 
& $ <60$ & $<6.43$  
& $ $\nodata$ $ & $ $\nodata$ $  
& $ $\nodata$ $ & $ $\nodata$ $  
\\
\sidehead{Mg II LTE line profile synthesis}
$1201$ & $4481.13$ & $4$ & $8.86$ & $0.73$ & & & \\[-0.8ex]
$1201$ & $4481.15$ & $4$ & $8.86$ & $-0.57$ 
& $ 240$ & $ 5.94$  
& $ 200$ & $ 6.10$  
& $ 200$ & $ 6.08$  \\[-0.8ex]
$1201$ & $4481.33$ & $4$ & $8.86$ & $0.57$ & & & 
\\
\sidehead{Si II LTE}
$1401$ & $3853.660$ & $1$ & $6.86$ & $-1.60$ 
& $  50$ & $ 6.48$  
& $  50$ & $ 6.48$  
& $  50$ & $ 6.51$  
\\
$1401$ & $3856.020$ & $1$ & $6.86$ & $-0.65$ 
& $ 150$ & $ 6.68$  
& $ 140$ & $ 6.74$  
& $ 140$ & $ 6.66$  
\\
$1401$ & $3862.590$ & $1$ & $6.86$ & $-0.90$ 
& $ 120$ & $ 6.59$  
& $ 115$ & $ 6.66$  
& $ 115$ & $ 6.60$  
\\
$1401$ & $4128.050$ & $3$ & $9.84$ & $0.31$ 
& $  80$ & $ 6.83$  
& $  70$ & $ 6.69$  
& $  70$ & $ 6.78$  
\\
$1401$ & $4130.890$ & $3$ & $9.84$ & $0.46$ 
& $  70$ & $ 6.56$  
& $  75$ & $ 6.62$  
& $  75$ & $ 6.70$  
\\
$1401$ & $5041.020$ & $5$ & $10.07$ & $0.17$ 
& $  35$ & $ 6.65$  
& $  40$ & $ 6.64$  
& $  40$ & $ 6.81$  
\\
$1401$ & $5055.980$ & $5$ & $10.07$ & $0.44$ 
& $  50$ & $ 6.65$  
& $  50$ & $ 6.57$  
& $  50$ & $ 6.71$  
\\
$1401$ & $6347.090$ & $2$ & $8.12$ & $0.23$ 
& $ 140$ & $ 6.76$  
& $ 125$ & $ 6.72$  
& $ 125$ & $ 6.68$  
\\
$1401$ & $6371.370$ & $2$ & $8.12$ & $-0.08$ 
& $  80$ & $ 6.45$  
& $  95$ & $ 6.67$  
& $  95$ & $ 6.65$  
\\
\sidehead{Sc II LTE}
$2101$ & $4246.820$ & $7$ & $0.32$ & $0.24$ 
& $ 140$ & $ 1.89$  
& $ 100$ & $ 1.87$  
& $ 140$ & $ 1.86$  
\\
$2101$ & $4314.080$ & $15$ & $0.62$ & $-0.10$ 
& $  75$ & $ 1.84$  
& $  40$ & $ 1.75$  
& $  80$ & $ 1.77$  
\\
$2101$ & $4320.730$ & $15$ & $0.61$ & $-0.21$ 
& $  80$ & $ 1.99$  
& $  40$ & $ 1.86$  
& $  80$ & $ 1.88$  
\\
$2101$ & $4325.000$ & $15$ & $0.60$ & $-0.44$ 
& $  50$ & $ 1.91$  
& $ $\nodata$ $ & $ $\nodata$ $  
& $  60$ & $ 1.89$  
\\
$2101$ & $4374.460$ & $14$ & $0.62$ & $-0.42$ 
& $  50$ & $ 1.90$  
& $  30$ & $ 1.91$  
& $  55$ & $ 1.83$  
\\
$2101$ & $4415.560$ & $14$ & $0.60$ & $-0.64$ 
& $  30$ & $ 1.83$  
& $ $\nodata$ $ & $ $\nodata$ $  
& $ $\nodata$ $ & $ $\nodata$ $  
\\
\sidehead{Ti II}
$2201$ & $3741.630$ & $72$ & $1.58$ & $-0.11$ 
& $ $\nodata$ $ & $ $\nodata$ $  
& $ 140$ & $ 4.08$  
& $ $\nodata$ $ & $ $\nodata$ $  
\\
$2201$ & $3900.550$ & $34$ & $1.13$ & $-0.44$ 
& $ $\nodata$ $ & $ $\nodata$ $  
& $ 130$ & $ 3.84$  
& $ $\nodata$ $ & $ $\nodata$ $  
\\
$2201$ & $3913.470$ & $34$ & $1.12$ & $-0.53$ 
& $ $\nodata$ $ & $ $\nodata$ $  
& $ 135$ & $ 3.99$  
& $ $\nodata$ $ & $ $\nodata$ $  
\\
$2201$ & $3932.020$ & $34$ & $1.13$ & $-1.78$ 
& $ $\nodata$ $ & $ $\nodata$ $  
& $  32$ & $ 3.99$  
& $ $\nodata$ $ & $ $\nodata$ $  
\\
$2201$ & $4012.400$ & $11$ & $0.57$ & $-1.61$ 
& $ 120$ & $ 4.10$  
& $  65$ & $ 3.84$  
& $ 110$ & $ 3.94$  
\\
$2201$ & $4025.130$ & $11$ & $0.61$ & $-1.98$ 
& $  60$ & $ 3.93$  
& $ $\nodata$ $ & $ $\nodata$ $  
& $  75$ & $ 3.98$  
\\
$2201$ & $4028.340$ & $87$ & $1.89$ & $-1.00$ 
& $  65$ & $ 3.94$  
& $  40$ & $ 3.86$  
& $  60$ & $ 3.79$  
\\
$2201$ & $4053.810$ & $87$ & $1.89$ & $-1.21$ 
& $  55$ & $ 4.04$  
& $  45$ & $ 4.13$  
& $  80$ & $ 4.21$  
\\
$2201$ & $4161.540$ & $21$ & $1.08$ & $-2.36$ 
& $  30$ & $ 4.25$  
& $ $\nodata$ $ & $ $\nodata$ $  
& $  30$ & $ 4.13$  
\\
$2201$ & $4163.630$ & $105$ & $2.59$ & $-0.40$ 
& $  80$ & $ 3.97$  
& $  60$ & $ 3.98$  
& $ 100$ & $ 4.11$  
\\
$2201$ & $4171.920$ & $105$ & $2.60$ & $-0.56$ 
& $  55$ & $ 3.89$  
& $  60$ & $ 4.15$  
& $ 100$ & $ 4.27$  
\\
$2201$ & $4290.220$ & $41$ & $1.16$ & $-1.12$ 
& $ 150$ & $ 4.31$  
& $  75$ & $ 3.84$  
& $ 110$ & $ 3.85$  
\\
$2201$ & $4294.090$ & $20$ & $1.08$ & $-1.11$ 
& $ 110$ & $ 3.84$  
& $  75$ & $ 3.77$  
& $ 120$ & $ 3.89$  
\\
$2201$ & $4300.060$ & $41$ & $1.18$ & $-0.77$ 
& $ 150$ & $ 3.98$  
& $ 125$ & $ 4.05$  
& $ 160$ & $ 4.14$  
\\
$2201$ & $4301.920$ & $41$ & $1.16$ & $-1.16$ 
& $  90$ & $ 3.77$  
& $  70$ & $ 3.83$  
& $ 100$ & $ 3.79$  
\\
$2201$ & $4307.900$ & $41$ & $1.16$ & $-1.29$ 
& $  95$ & $ 3.94$  
& $  80$ & $ 4.06$  
& $ 105$ & $ 3.97$  
\\
$2201$ & $4312.860$ & $41$ & $1.18$ & $-1.16$ 
& $  90$ & $ 3.78$  
& $  70$ & $ 3.84$  
& $ 100$ & $ 3.80$  
\\
$2201$ & $4314.970$ & $41$ & $1.16$ & $-1.13$ 
& $  90$ & $ 3.74$  
& $  80$ & $ 3.90$  
& $ 120$ & $ 3.97$  
\\
$2201$ & $4316.800$ & $94$ & $2.05$ & $-1.42$ 
& $  30$ & $ 4.00$  
& $ $\nodata$ $ & $ $\nodata$ $  
& $  40$ & $ 4.05$  
\\
$2201$ & $4320.960$ & $41$ & $1.16$ & $-1.87$ 
& $  60$ & $ 4.19$  
& $  40$ & $ 4.18$  
& $  40$ & $ 3.84$  
\\
$2201$ & $4330.700$ & $41$ & $1.18$ & $-2.04$ 
& $  30$ & $ 3.99$  
& $ $\nodata$ $ & $ $\nodata$ $  
& $  60$ & $ 4.27$  
\\
$2201$ & $4374.820$ & $93$ & $2.06$ & $-1.29$ 
& $  40$ & $ 4.03$  
& $  30$ & $ 4.08$  
& $  60$ & $ 4.17$  
\\
$2201$ & $4386.840$ & $104$ & $2.60$ & $-1.26$ 
& $  25$ & $ 4.14$  
& $ $\nodata$ $ & $ $\nodata$ $  
& $ $\nodata$ $ & $ $\nodata$ $  
\\
$2201$ & $4394.050$ & $51$ & $1.23$ & $-1.59$ 
& $ $\nodata$ $ & $ $\nodata$ $  
& $  40$ & $ 3.94$  
& $ $\nodata$ $ & $ $\nodata$ $  
\\
$2201$ & $4395.030$ & $19$ & $1.08$ & $-0.66$ 
& $ 150$ & $ 3.78$  
& $ 140$ & $ 4.05$  
& $ $\nodata$ $ & $ $\nodata$ $  
\\
$2201$ & $4395.850$ & $61$ & $1.24$ & $-2.17$ 
& $ $\nodata$ $ & $ $\nodata$ $  
& $  20$ & $ 4.17$  
& $  45$ & $ 4.26$  
\\
$2201$ & $4399.770$ & $51$ & $1.24$ & $-1.27$ 
& $ 100$ & $ 4.02$  
& $  70$ & $ 3.98$  
& $ 130$ & $ 4.27$  
\\
$2201$ & $4417.720$ & $40$ & $1.15$ & $-1.43$ 
& $ 100$ & $ 4.11$  
& $  55$ & $ 3.91$  
& $ 115$ & $ 4.19$  
\\
$2201$ & $4421.940$ & $93$ & $2.06$ & $-1.39$ 
& $  30$ & $ 3.97$  
& $ $\nodata$ $ & $ $\nodata$ $  
& $ $\nodata$ $ & $ $\nodata$ $  
\\
$2201$ & $4443.800$ & $19$ & $1.08$ & $-0.70$ 
& $ 160$ & $ 3.92$  
& $ 130$ & $ 3.95$  
& $ 170$ & $ 4.11$  
\\
$2201$ & $4464.450$ & $40$ & $1.16$ & $-2.08$ 
& $  40$ & $ 4.15$  
& $  25$ & $ 4.13$  
& $  60$ & $ 4.29$  
\\
$2201$ & $4468.490$ & $31$ & $1.13$ & $-0.62$ 
& $ 160$ & $ 3.87$  
& $ 130$ & $ 3.90$  
& $ $\nodata$ $ & $ $\nodata$ $  
\\
$2201$ & $4549.620$ & $82$ & $1.58$ & $-0.47$ 
& $ 150$ & $ 3.93$  
& $ $\nodata$ $ & $ $\nodata$ $  
& $ $\nodata$ $ & $ $\nodata$ $  
\\
$2201$ & $4563.760$ & $50$ & $1.22$ & $-0.96$ 
& $ 120$ & $ 3.86$  
& $ $\nodata$ $ & $ $\nodata$ $  
& $ $\nodata$ $ & $ $\nodata$ $  
\\
$2201$ & $4571.960$ & $82$ & $1.57$ & $-0.52$ 
& $ 120$ & $ 3.67$  
& $ $\nodata$ $ & $ $\nodata$ $  
& $ $\nodata$ $ & $ $\nodata$ $  
\\
$2201$ & $4805.090$ & $92$ & $2.06$ & $-1.12$ 
& $  60$ & $ 4.06$  
& $  30$ & $ 3.88$  
& $  50$ & $ 3.86$  
\\
$2201$ & $4874.010$ & $114$ & $3.09$ & $-0.79$ 
& $  30$ & $ 4.09$  
& $ $\nodata$ $ & $ $\nodata$ $  
& $  45$ & $ 4.22$  
\\
$2201$ & $4911.190$ & $114$ & $3.12$ & $-0.33$ 
& $  45$ & $ 3.87$  
& $  35$ & $ 3.91$  
& $  65$ & $ 4.02$  
\\
$2201$ & $5129.150$ & $86$ & $1.89$ & $-1.40$ 
& $  40$ & $ 3.97$  
& $ $\nodata$ $ & $ $\nodata$ $  
& $  40$ & $ 3.87$  
\\
$2201$ & $5185.910$ & $86$ & $1.89$ & $-1.35$ 
& $  35$ & $ 3.85$  
& $ $\nodata$ $ & $ $\nodata$ $  
& $  40$ & $ 3.82$  
\\
$2201$ & $5188.680$ & $70$ & $1.58$ & $-1.22$ 
& $  45$ & $ 3.63$  
& $  40$ & $ 3.77$  
& $  70$ & $ 3.80$  
\\
$2201$ & $5226.540$ & $70$ & $1.57$ & $-1.29$ 
& $  50$ & $ 3.75$  
& $  60$ & $ 4.08$  
& $  80$ & $ 3.96$  
\\
$2201$ & $5418.750$ & $69$ & $1.58$ & $-2.00$ 
& $  20$ & $ 3.98$  
& $ $\nodata$ $ & $ $\nodata$ $  
& $ $\nodata$ $ & $ $\nodata$ $  
\\
\sidehead{Cr I LTE}
$2400$ & $4254.330$ & $1$ & $0.00$ & $-0.11$ 
& $  40$ & $ 4.71$  
& $  30$ & $ 5.06$  
& $  45$ & $ 4.56$  
\\
$2400$ & $4274.800$ & $1$ & $0.00$ & $-0.23$ 
& $  20$ & $ 4.47$  
& $ $\nodata$ $ & $ $\nodata$ $  
& $ $\nodata$ $ & $ $\nodata$ $  
\\
$2400$ & $5208.420$ & $7$ & $0.94$ & $0.16$ 
& $  20$ & $ 4.71$  
& $ $\nodata$ $ & $ $\nodata$ $  
& $ $\nodata$ $ & $ $\nodata$ $  
\\
\sidehead{Cr II LTE}
$2401$ & $4051.930$ & $19$ & $3.10$ & $-2.19$ 
& $  30$ & $ 4.68$  
& $ $\nodata$ $ & $ $\nodata$ $  
& $  40$ & $ 4.76$  
\\
$2401$ & $4111.000$ & $0$ & $3.74$ & $-1.92$ 
& $  25$ & $ 4.77$  
& $  25$ & $ 4.88$  
& $  20$ & $ 4.58$  
\\
$2401$ & $4224.860$ & $162$ & $5.33$ & $-1.06$ 
& $  25$ & $ 4.99$  
& $ $\nodata$ $ & $ $\nodata$ $  
& $ $\nodata$ $ & $ $\nodata$ $  
\\
$2401$ & $4242.360$ & $31$ & $3.87$ & $-1.17$ 
& $  80$ & $ 4.78$  
& $  60$ & $ 4.72$  
& $  70$ & $ 4.64$  
\\
$2401$ & $4252.630$ & $31$ & $3.86$ & $-2.02$ 
& $  25$ & $ 4.94$  
& $ $\nodata$ $ & $ $\nodata$ $  
& $  20$ & $ 4.76$  
\\
$2401$ & $4261.910$ & $31$ & $3.86$ & $-1.53$ 
& $  55$ & $ 4.88$  
& $  45$ & $ 4.89$  
& $  35$ & $ 4.56$  
\\
$2401$ & $4275.570$ & $31$ & $3.86$ & $-1.71$ 
& $  40$ & $ 4.88$  
& $  35$ & $ 4.92$  
& $  40$ & $ 4.81$  
\\
$2401$ & $4284.190$ & $31$ & $3.85$ & $-1.86$ 
& $  25$ & $ 4.77$  
& $  25$ & $ 4.89$  
& $  30$ & $ 4.80$  
\\
$2401$ & $4558.650$ & $44$ & $4.07$ & $-0.66$ 
& $ 120$ & $ 4.77$  
& $ $\nodata$ $ & $ $\nodata$ $  
& $ $\nodata$ $ & $ $\nodata$ $  
\\
$2401$ & $4588.200$ & $44$ & $4.07$ & $-0.64$ 
& $ 110$ & $ 4.66$  
& $ $\nodata$ $ & $ $\nodata$ $  
& $ $\nodata$ $ & $ $\nodata$ $  
\\
$2401$ & $4618.800$ & $44$ & $4.07$ & $-1.11$ 
& $  60$ & $ 4.65$  
& $ $\nodata$ $ & $ $\nodata$ $  
& $ $\nodata$ $ & $ $\nodata$ $  
\\
$2401$ & $4824.130$ & $30$ & $3.87$ & $-1.22$ 
& $  70$ & $ 4.71$  
& $  80$ & $ 4.97$  
& $  70$ & $ 4.66$  
\\
$2401$ & $4848.240$ & $30$ & $3.86$ & $-1.14$ 
& $  65$ & $ 4.57$  
& $  80$ & $ 4.88$  
& $  75$ & $ 4.62$  
\\
$2401$ & $4876.410$ & $30$ & $3.86$ & $-1.46$ 
& $  55$ & $ 4.79$  
& $  50$ & $ 4.86$  
& $  40$ & $ 4.54$  
\\
$2401$ & $4884.610$ & $30$ & $3.86$ & $-2.13$ 
& $  15$ & $ 4.78$  
& $ $\nodata$ $ & $ $\nodata$ $  
& $ $\nodata$ $ & $ $\nodata$ $  
\\
$2401$ & $5237.330$ & $43$ & $4.06$ & $-1.16$ 
& $  55$ & $ 4.62$  
& $ $\nodata$ $ & $ $\nodata$ $  
& $  50$ & $ 4.50$  
\\
$2401$ & $5274.960$ & $43$ & $4.05$ & $-1.29$ 
& $  45$ & $ 4.63$  
& $ $\nodata$ $ & $ $\nodata$ $  
& $  40$ & $ 4.50$  
\\
$2401$ & $5334.870$ & $43$ & $4.07$ & $-1.56$ 
& $  25$ & $ 4.60$  
& $ $\nodata$ $ & $ $\nodata$ $  
& $  40$ & $ 4.78$  
\\
\sidehead{Fe I LTE}
$2600$ & $3787.880$ & $21$ & $1.01$ & $-0.84$ 
& $  35$ & $ 6.03$  
& $ $\nodata$ $ & $ $\nodata$ $  
& $  60$ & $ 6.15$  
\\
$2600$ & $3820.430$ & $20$ & $0.86$ & $0.16$ 
& $ 150$ & $ 6.18$  
& $ 110$ & $ 6.32$  
& $ 150$ & $ 6.15$  
\\
$2600$ & $3825.880$ & $20$ & $0.91$ & $-0.03$ 
& $ 110$ & $ 5.95$  
& $  90$ & $ 6.30$  
& $ 135$ & $ 6.14$  
\\
$2600$ & $3859.910$ & $4$ & $0.00$ & $-0.71$ 
& $ 120$ & $ 6.05$  
& $  90$ & $ 6.33$  
& $ 140$ & $ 6.21$  
\\
$2600$ & $3920.260$ & $4$ & $0.12$ & $-1.75$ 
& $  30$ & $ 6.19$  
& $ $\nodata$ $ & $ $\nodata$ $  
& $ $\nodata$ $ & $ $\nodata$ $  
\\
$2600$ & $3922.910$ & $4$ & $0.05$ & $-1.65$ 
& $  50$ & $ 6.32$  
& $ $\nodata$ $ & $ $\nodata$ $  
& $ $\nodata$ $ & $ $\nodata$ $  
\\
$2600$ & $3927.920$ & $4$ & $0.11$ & $-1.59$ 
& $  30$ & $ 6.02$  
& $ $\nodata$ $ & $ $\nodata$ $  
& $  50$ & $ 6.09$  
\\
$2600$ & $4005.240$ & $43$ & $1.56$ & $-0.61$ 
& $  50$ & $ 6.37$  
& $ $\nodata$ $ & $ $\nodata$ $  
& $  55$ & $ 6.23$  
\\
$2600$ & $4045.810$ & $43$ & $1.48$ & $0.28$ 
& $ 130$ & $ 6.22$  
& $  85$ & $ 6.30$  
& $ 130$ & $ 6.13$  
\\
$2600$ & $4063.600$ & $43$ & $1.56$ & $0.06$ 
& $  85$ & $ 6.05$  
& $  60$ & $ 6.30$  
& $ 110$ & $ 6.15$  
\\
$2600$ & $4071.740$ & $43$ & $1.61$ & $-0.02$ 
& $ $\nodata$ $ & $ $\nodata$ $  
& $  55$ & $ 6.36$  
& $ 100$ & $ 6.16$  
\\
$2600$ & $4132.060$ & $43$ & $1.61$ & $-0.67$ 
& $  30$ & $ 6.17$  
& $ $\nodata$ $ & $ $\nodata$ $  
& $  40$ & $ 6.12$  
\\
$2600$ & $4143.870$ & $43$ & $1.56$ & $-0.51$ 
& $  45$ & $ 6.19$  
& $  30$ & $ 6.46$  
& $  55$ & $ 6.12$  
\\
$2600$ & $4187.040$ & $152$ & $2.45$ & $-0.55$ 
& $  15$ & $ 6.31$  
& $ $\nodata$ $ & $ $\nodata$ $  
& $ $\nodata$ $ & $ $\nodata$ $  
\\
$2600$ & $4187.800$ & $152$ & $2.43$ & $-0.55$ 
& $  15$ & $ 6.29$  
& $ $\nodata$ $ & $ $\nodata$ $  
& $ $\nodata$ $ & $ $\nodata$ $  
\\
$2600$ & $4202.030$ & $42$ & $1.48$ & $-0.71$ 
& $  45$ & $ 6.33$  
& $ $\nodata$ $ & $ $\nodata$ $  
& $  50$ & $ 6.19$  
\\
\sidehead{Fe II LTE}
$2601$ & $3783.350$ & $14$ & $2.27$ & $-3.16$ 
& $  80$ & $ 6.37$  
& $  85$ & $ 6.60$  
& $  85$ & $ 6.38$  
\\
$2601$ & $4024.550$ & $127$ & $4.49$ & $-2.48$ 
& $  35$ & $ 6.67$  
& $  30$ & $ 6.65$  
& $ $\nodata$ $ & $ $\nodata$ $  
\\
$2601$ & $4122.640$ & $28$ & $2.58$ & $-3.38$ 
& $  55$ & $ 6.51$  
& $ $\nodata$ $ & $ $\nodata$ $  
& $ $\nodata$ $ & $ $\nodata$ $  
\\
$2601$ & $4173.450$ & $27$ & $2.58$ & $-2.18$ 
& $ 160$ & $ 6.40$  
& $ 155$ & $ 6.70$  
& $ 150$ & $ 6.35$  
\\
$2601$ & $4178.860$ & $28$ & $2.58$ & $-2.48$ 
& $ 140$ & $ 6.46$  
& $ 130$ & $ 6.63$  
& $ 130$ & $ 6.38$  
\\
$2601$ & $4258.150$ & $28$ & $2.70$ & $-3.40$ 
& $  50$ & $ 6.55$  
& $ $\nodata$ $ & $ $\nodata$ $  
& $  40$ & $ 6.35$  
\\
$2601$ & $4273.320$ & $27$ & $2.70$ & $-3.34$ 
& $  35$ & $ 6.29$  
& $  50$ & $ 6.62$  
& $  45$ & $ 6.36$  
\\
$2601$ & $4351.770$ & $27$ & $2.70$ & $-2.10$ 
& $ 160$ & $ 6.37$  
& $ 150$ & $ 6.59$  
& $ 140$ & $ 6.19$  
\\
$2601$ & $4369.410$ & $28$ & $2.78$ & $-3.66$ 
& $ $\nodata$ $ & $ $\nodata$ $  
& $  30$ & $ 6.69$  
& $ $\nodata$ $ & $ $\nodata$ $  
\\
$2601$ & $4416.830$ & $27$ & $2.78$ & $-2.61$ 
& $ 120$ & $ 6.49$  
& $ 110$ & $ 6.61$  
& $ 110$ & $ 6.38$  
\\
$2601$ & $4491.400$ & $37$ & $2.86$ & $-2.69$ 
& $  80$ & $ 6.25$  
& $ 100$ & $ 6.63$  
& $ $\nodata$ $ & $ $\nodata$ $  
\\
$2601$ & $5197.570$ & $49$ & $3.23$ & $-2.10$ 
& $ 125$ & $ 6.33$  
& $ 125$ & $ 6.57$  
& $ 130$ & $ 6.40$  
\\
$2601$ & $5234.620$ & $49$ & $3.22$ & $-2.05$ 
& $ 140$ & $ 6.42$  
& $ $\nodata$ $ & $ $\nodata$ $  
& $ 130$ & $ 6.34$  
\\
$2601$ & $5276.000$ & $44$ & $3.20$ & $-1.95$ 
& $ 135$ & $ 6.25$  
& $ 145$ & $ 6.65$  
& $ 145$ & $ 6.41$  
\\
$2601$ & $5316.620$ & $49$ & $3.15$ & $-1.85$ 
& $ $\nodata$ $ & $ $\nodata$ $  
& $ 150$ & $ 6.58$  
& $ 150$ & $ 6.33$  
\\
$2601$ & $6247.560$ & $74$ & $3.89$ & $-2.36$ 
& $  50$ & $ 6.35$  
& $  65$ & $ 6.64$  
& $  55$ & $ 6.35$  
\\
$2601$ & $6416.910$ & $74$ & $3.89$ & $-2.70$ 
& $ $\nodata$ $ & $ $\nodata$ $  
& $  45$ & $ 6.74$  
& $ $\nodata$ $ & $ $\nodata$ $  
\\
\enddata
\end{deluxetable}

\clearpage 

\begin{deluxetable}{ccccccccccccc}
\tabletypesize{\footnotesize}
\tablecaption{\label{SexA-Mg_corr} Magnesium NLTE Corrections}
\tablewidth{0pt}
\tablecolumns{13}
\tablehead{
\colhead{$\lambda$}               &  
\colhead{ }                       & 
\colhead{$\chi$}                  & 
\colhead{ }                       & 
\multicolumn{3}{c}{SexA-754}      & 
\multicolumn{3}{c}{SexA-1344}     & 
\multicolumn{3}{c}{SexA-1911}     \\
\colhead{[\AA]}                   & 
\colhead{Levels\tablenotemark{a}} & 
\colhead{[eV]}                    & 
\colhead{$\log(gf)$}              & 
\colhead{$W_{\lambda}$}           & 
\colhead{LTE}                     & 
\colhead{NLTE}                    & 
\colhead{$W_{\lambda}$}           & 
\colhead{LTE}                     & 
\colhead{NLTE}                    & 
\colhead{$W_{\lambda}$}           & 
\colhead{LTE}                     & 
\colhead{NLTE}              
}
\startdata
\sidehead{\ion{Mg}{1}} 
3829.36 & $3p^3P^0-3d^3D$ & 2.71 & $-0.21$ & 
        140 & 6.47  & 6.50  &  
         80 & 6.39  & 6.49  &
        140 & 6.39  & 6.42  \\
5167.32 & $3p^3P^0-4s^3S$ & 2.71 & $-0.86$ &  
         75 & 6.40  & 6.49  &  
         35 & 6.41  & 6.54  &
         85 & 6.33  & 6.40  \\
5172.68 & $3p^3P^0-4s^3S$ & 2.71 & $-0.38$ & 
        135 & 6.46  & 6.50  & 
         80 & 6.45  & 6.58  &
        150 & 6.52  & 6.50  \\
5183.60 & $3p^3P^0-4s^3S$ & 2.72 & $-0.16$ & 
        160 & 6.50  & 6.49  & 
        100 & 6.44  & 6.55  &
        160 & 6.43  & 6.42  \\
\sidehead{\ion{Mg}{2}} 
3848.21 & $3d^2D-5p^2P^0$ & 8.86 & $-1.56$ & 
        $<20$ & $<6.84$  & $<6.84$ &
        \nodata &   \nodata &   \nodata &
        \nodata &   \nodata &   \nodata \\
4390.57 & $4p^2P^0-5d^2D$ &10.00 & $-0.50$ &  
        $<20$ & $<6.52$  & $<6.52$  &
        \nodata & \nodata   & \nodata   &
        \nodata & \nodata   & \nodata\\
%{\it 4481.13}\tablenotemark{b} 
%        & $3d^2D-4f^2F^0$ & {\it 8.86}  & {\it 0.73} & & & & \\
%{\it 4481.15}\tablenotemark{b} 
%        & $3d^2D-4f^2F^0$ & {\it 8.86}  & {\it -0.57} & 
%        {\it 240} & {\it 6.48}  & \nodata  &
%        {\it 200} & {\it 6.59}  & \nodata  &
%        {\it 200} & {\it 6.56}  & \nodata  \\
%{\it 4481.33}\tablenotemark{b} 
%        & $3d^2D-4f^2F^0$ & {\it 8.86}  & {\it 0.57} & & & & \\
7877.05 & $4p^2P^0-4d^2D$ &10.00 & $ 0.39$ & 
        $<60$ & $<6.80$  & $<6.69$  &
        \nodata & \nodata   & \nodata   &
        \nodata & \nodata   & \nodata\\
7896.37 & $4p^2P^0-4d^2D$ &10.00 & $ 0.65$ & 
        $<60$ & $<6.54$  & $<6.43$  &
        \nodata & \nodata   & \nodata   &
        \nodata & \nodata   & \nodata  \\
\enddata
\tablenotetext{a}{The levels are described with their configurations
  and terms: configuration $n[l]$, term $ ^{2S+1}[L]$, the superscript
  $^0$ means the parity is odd and the absence of subscript means the
  parity is even.}
%\tablenotetext{b}{Not used in any of the analyses (cf. text)}
%
\end{deluxetable}

%
%%%%%%%%%%%%%%%%%%%%%%%%%%%%%%%%%%%%%%%%%%%%%%%%%%%%%%%%%%%%%%%%%%%%%%%%%%%
%
\end{document}